\documentclass{aa}

\usepackage{natbib,twoopt}
\usepackage[breaklinks=true,colorlinks=true,citecolor=blue,linkcolor=blue,urlcolor=blue, backref=false]{hyperref}
\usepackage{url}
\usepackage{mathrsfs}

\usepackage{txfonts}
\usepackage{multirow}
\usepackage{colortbl}
\usepackage{xcolor}
\usepackage{orcidlink}
\usepackage{academicons}
\usepackage{textcomp}
\usepackage[capitalise]{cleveref}
\usepackage[normalem]{ulem}
\bibpunct{(}{)}{;}{a}{}{,} 
\usepackage{bm}
\usepackage{algpseudocode}
\usepackage{algorithm}

\newcommand{\Om}{$\Omega_{\rm m}$}
\newcommand{\sigmaeight}{$\sigma_8$}

\newcommand{\lcdm}{$\mathrm{\Lambda CDM}$}
\newcommand{\Mpch}{$h^{-1}\,\mbox{Mpc}$}

\newcommand{\MpchInv}{$\mbox{Mpc}/h$}

\definecolor{RoyalBlue}{rgb}{0.25,.41,.88}
\definecolor{airforceblue}{rgb}{0.36, 0.54, 0.66}

\newcommand{\eg}{\textit{e}.\textit{g}.}

\newcommand{\erw}{\mathbb{E}}
\DeclareMathOperator{\KL}{KL}
\newcommand{\calD}{\mathcal{D}}
\newcommand{\calN}{\mathcal{N}}
\DeclareMathOperator{\ELBO}{ELBO}

\begin{document} 

   \title{Bayesian deep learning  for  cosmic volumes with modified gravity}
   \titlerunning{Garc\'ia-Farieta,  Hort\'ua \& Kitaura}
    \author{J.E. Garc\'ia-Farieta, 
    \inst{1,2}\orcidlink{0000-0001-6667-5471}\thanks{jorge.farieta@iac.es}, 
          H\'ector J Hort\'ua\inst{3,4} \orcidlink{0000-0002-3396-2404}\thanks{hhortuao@unbosque.edu.co} \and 
          Francisco-Shu Kitaura\inst{1,2} \orcidlink{0000-0002-9994-759X} 
          }
    
    \institute{Instituto de Astrof\'{\i}sica de Canarias, s/n, E-38205, La Laguna, Tenerife, Spain \email{jorge.farieta@iac.es}
    \and
    Departamento de Astrof\'{\i}sica, Universidad de La Laguna, E-38206, La Laguna, Tenerife, Spain 
    \and
    Grupo Signos, Departamento de Matemáticas, Universidad El Bosque, Bogotá, Colombia
    \and
    Maestría en Ciencia de Datos, Universidad Escuela Colombiana de Ingeniería Julio Garavito, Bogotá, Colombia
    }

   \authorrunning{Author}
   \date{Received September X, XXXX; accepted March X, XXXX}

% \abstract{}{}{}{}{} 
% 5 {} token are mandatory
 
  \abstract
   {The new generation of galaxy surveys will provide unprecedented data that will allow us to test gravity deviations at cosmological scales at a much higher precision than could be achieved previously. A robust cosmological analysis of the large-scale structure demands exploiting the nonlinear information encoded in the cosmic web. Machine-learning techniques provide these tools, but no a priori assessment of the uncertainties.}
   {We extract cosmological parameters from modified gravity (MG) simulations through deep neural networks that include uncertainty estimations.}
   {We implemented Bayesian neural networks (BNNs) with an enriched approximate posterior distribution considering two cases: the first case with a single Bayesian last layer (BLL), and the other case with Bayesian layers at all levels (FullB). We trained both BNNs with real-space density fields and power spectra from a suite of 2000 dark matter-only particle-mesh $N$-body simulations including MG models relying on MG-PICOLA, covering 256 $h^{-1}$ Mpc side cubical volumes with 128$^3$ particles.} 
  {BNNs excel in accurately predicting parameters for \Om\ and \sigmaeight\ and their respective correlation with the MG parameter. Furthermore, we find that BNNs yield well-calibrated uncertainty estimates that overcome the over- and under-estimation issues in traditional neural networks. The MG parameter leads to a significant degeneracy, and \sigmaeight\ might be one possible explanation of the poor MG predictions. Ignoring MG, we obtain a deviation of the relative errors in \Om\ and \sigmaeight\ by $30\%$ at least. Moreover, we report consistent results from the density field and power spectrum analysis and comparable results between BLL and FullB experiments. This halved the computing time. This work contributes to preparing the path for extracting cosmological parameters from complete small cosmic volumes towards the highly nonlinear regime.}
   {}

   \keywords{cosmology: – large-scale structure of Universe - cosmological parameters; methods: data analysis - statistical - numerical
               }

   \maketitle
%
%-------------------------------------------------------------------
\section{Introduction}\label{sec:intro}
Cosmic acceleration is one of the most critical concerns in modern cosmology. In the context of the concordance model \lcdm\ ($\Lambda$-cold dark matter), this acceleration is attributed to the existence of a fluid with negative pressure that is represented by the cosmological constant, $\Lambda$, in General Relativity (GR) equations. However, this fluid introduces some conceptual and theoretical issues that have not been fully addressed, either observationally or theoretically. Alternative theories, such as MG models, have attracted attention as a natural explanation for cosmic acceleration without invoking a cosmological constant \citep[see, e.g.,][for a recent review]{Nojiri_MGrev2017PhR}. Among the plethora of alternative models, some parametrisations of $f(R)$ gravity have gained popularity because they can accurately reproduce the standard model predictions. The two cosmological scenarios \lcdm\ and $f(R)$ are indeed highly successful in providing an accurate description of the Universe on large scales, from cosmic microwave background (CMB) observations to the data of galaxy clustering \citep{Berti_testGRMG_2015}.\\

Unlike the standard scenario, the $f(R)$ models do not require a cosmological constant, but instead modify the behaviour of gravity itself. The modification of Einstein's GR involves the addition of a scalar field that emulates cosmic acceleration. This feature of $f(R)$ models has made them perfect templates for tracking departures from standard gravity. Consequently, a crucial task within the scope of precision cosmology is to quantify the potential variations in gravity using appropriate techniques that are sensitive to MG effects. Some of the approaches to achieve this aim include using clustering anisotropies \citep{2012MNRAS.425.2128J, Garfa2019-MGMnu, Aguayo2019, 2020MNRAS.494.1658G}, tracer bias and sample selection \citep{2021PhRvD.103j3524G}, cosmic voids \citep{2017PhRvD..95b4018V, 2019A&A...632A..52P, 2021MNRAS.504.5021C}, halo mass functions \citep{2019MNRAS.486.3927H, 2022PhRvD.105d3538G}, and peculiar velocities \citep{2016MNRAS.458.2725J, 2016A&A...595A..40I,2023MNRAS.518.5929L}.\\ 

The matter distribution is a rich source of cosmological information that has been exploited for many years through various techniques. One of the most frequently used techniques to extract information from the large-scale structure (LSS) data relies on the two-point statistics as described by the two-point correlation function or its equivalent in Fourier space, the matter power spectrum. Despite its success in capturing all possible cosmological information contained in a density field, it fails to capture features affected by the non-Gaussian nature of density perturbations, and its accuracy and precision cannot be relied upon for probing small angular scales. Since the estimators up to second order do not contain all cosmological information, other techniques beyond the two-point statistics have been studied to extract additional information, such as $N$-point correlation functions \citep{2001ASPC..252..201P,2003MNRAS.340..580T,2022MNRAS.515.6133Z,2022A&A...667A.129B,2022JCAP...09..033V,2022MNRAS.509.2457P}, Minkowski functionals \citep{2012PhRvD..85j3513K,2003PASJ...55..911H,2017PhRvL.118r1301F}, peak count statistics \citep{2016MNRAS.463.3653K, 2017A&A...599A..79P,2018JCAP...10..051F,2021MNRAS.506.1623H}, density split statistics \citep{Paillas2021MNRAS}, cosmic shear \citep{2015RPPh...78h6901K,2001A&A...374..757V}, cosmic voids \cite{2015MNRAS.451.1036C,2012MNRAS.426..440B,2016PhRvL.117i1302H,2010MNRAS.403.1392L}, and tomographic analyses based on the Alcock-Paczynski test \citep{2010ApJ...715L.185P,2019ApJ...878..137Z,2015MNRAS.450..807L,2019ApJ...887..125L,2023ApJ...953...98D}, and Bayesian inference~\citep{2023arXiv230201331N,2023JCAP...07..063K}. For an overview of contemporary cosmological probes, we refer to \cite{Weinberg_CosmicAcel_2013PhR} and \cite{2022LRR....25....6M}.\\

Recently,  deep  neural networks (DNNs)  have been proposed as a new alternative for not only recollecting the three-dimensional (3D) density field information without specifying the summary statistic beforehand, such as the power spectrum, but also for managing the demanding computational needs in astrophysics~\citep{220308056}. The method involving parameter inference and model comparison directly from simulations, that is, using high-dimensional data, is formally termed simulation-based inference \citep{2016arXiv161110242T, Lemos_2023a, Lemos_2023b, Hahn:2022zxa}. This technique is sometimes referred to as approximate Bayesian computation, implicit likelihood inference, or non-likelihood inference \citep{csillery2010approximate,10.1371/journal.pcbi.1002803,Beaumontannurev-ecolsys-102209-144621}. The CNN algorithms have been explored as a valuable tool in MG scenarios, mainly with applications in weak-lensing maps such as emulator building \citep{2021MNRAS.506.3049T} and also to investigate observational degeneracies between MG models and massive neutrinos \citep{2019MNRAS.487..104M,2019PhRvD.100b3508P}, CMB patch map analysis~\citep{Hort_a_2020},  and
N-body simulations~\citep{2021,oliveira2020fast,2020}. Bayesian neural networks (BNNs) have also been employed to identify and classify power spectra that deviate from \lcdm\ such as MG models, however \citep{2022PhRvD.105b3531M}.

Even though they can extract information from complex data, standard DNNs still overfit or memorise noisy labels during the training phase, and their point estimates are not always reliable. BNNs are extensions of DNNs that provide probabilistic properties on their outcomes and yield predictive uncertainties~\cite{10.5555/2986766.2986882,pmlr-v48-gal16}.  Some BBN approaches include stochastic MCMC~\cite{deng2019bayesian}, Bayes by backprop~\cite{10.5555/3045118.3045290}, deep ensembles~\cite{10.5555/3295222.3295387}, or Monte Carlo dropout and flipout~\cite{Gal2016Uncertainty,wen2018flipout}.
Other BNN alternatives employ  variational inference (VI) to infer posterior distributions for the network weights that are suitable to capture its uncertainties, which are propagated to the network output~\citep{NIPS2011_7eb3c8be,gunapati_jain_srijith_desai_2022}.
Although VI speeds up the computation of the posterior distribution when analytic approaches are considered,  these assumptions can also introduce a bias~\citep{2006.01490} that yields overconfident uncertainty predictions and significant deviations from the true posterior. ~\citet{Hort_a_2020a} and~\citet{2112.11865} added normalising flows on top of  BNNs to give the joint parameter distribution more flexibility. However, this approach was not implemented into the Bayesian framework, so that the bias is still preserved. In a recent work~\citep[][]{frontiershector}, the authors improved the previous method by applying
multiplicative normalising flows, which resulted in accurate uncertainty estimates. In this paper, we follow the same approach by building BNN models adapted to both 3D density field and its power spectra to constrain MG from cosmological simulations. 
We show that for non-Gaussian signals alone, it is possible to improve the posterior distributions, and when the additional information from the power spectrum is considered, they yield more significant performance improvements without underestimating the posterior distributions. The analysis focuses on simulated dark matter fields from cosmological models beyond \lcdm\ to assess whether the inference schemes can reliably capture as much information as possible from the nonlinear regime. However, the applicability of the proposed model can be adapted to be easily used with small-volume mock galaxy catalogues. This paper is organised as follows. Section 2 offers a summary of structure formation in MG cosmologies and the reference simulations we created to train and test the BNNs. In section 3 we briefly introduce the BNN concept, and section 4 shows the architectures and configuration we used in the paper. The results are presented in section 5, and an extended discussion of the findings is presented in section 6. Our conclusions are given in section 7.

%-------------------------------------------------------------------
\section{Large-scale structure in modified gravity}
In this section, we present the gravity model that coincides with $\Lambda$CDM in the limiting case of a vanishing $f(R)$ parameter introduced below.
\subsection{Structure formation and background}
In $f(R)$ cosmologies, the dynamics of matter is determined by the modified Einstein field equations. The most straightforward modification of GR that circumvents $\Lambda$ emerges by including a function of the curvature scalar in the Einstein-Hilbert action. In this modification, the equations of motion are enriched with a term that depends on the curvature scalar and that creates the same effect as dark energy \citep[for a review on different MG models see e.g.][]{Tsujikawa_Geff_2008, DeFelice_review_fR_2010}. For consistency across various cosmological scales, \citet[][hereafter HS]{HuSawicki_2007} proposed a viable $f(R)$ function that can satisfy tight constraints at Solar System scales and that also accurately describes the dynamics of the \lcdm\ background. In these models, the modified Einstein-Hilbert action is given by
\begin{equation}
S_{\rm EH}=\int \mathrm{d}^4 x \sqrt{-g}\left[\frac{R+f(R)}{16 \pi G}\right] \,,
\end{equation}
where $g$ is the metric tensor, $G$ is Newton's gravitational constant, $R$ is the curvature scalar, and $f(R)$ is a scalar function that constrains the additional degree of freedom. In the HS model, this function takes the form
\begin{equation}
f(R)=-m^2 \frac{c_1\left(-R / m^2\right)^n}{c_2\left(-R / m^2\right)^n+1}\,,
\end{equation}
where $n$, $c_1$, and $c_2$ are model parameters, and $m^2\equiv\Omega_{\rm m}H_0^2$, with \Om\ being the current fractional matter density, and
$H_0$ is the Hubble parameter at the present time. For $n=1$, which is the $f(R)$ model we consider in this paper, the function can be written as follows:
\begin{equation}
f(R)\approx-2 \Lambda+|f_{R0}| \frac{R_0^2}{R}\,.
\end{equation}
Here, $f_{R0}$ represents the dimensionless scalar field at the present time, meaning the only additional degree of freedom that stems from the derivative of $f(R)$ with respect to the curvature scalar, $f_R$. The modified Einstein field equations and analogous Friedmann equations that describe the HS model background can be obtained from minimising the action \citep[for a thorough derivation see][]{2007PhRvD..75d4004S}. To further understand the formation and evolution of large-scale structures in MG, it is crucial to describe the matter perturbations, $\delta_m$, around the background \citep[see][]{2007PhRvD..75d4004S}. The MG effects are captured by the growth of density perturbations in the matter-dominated era when the mildly nonlinear regime is important \citep{2008PhRvD..77b4048L}. In particular, when considering linear perturbations, the equations of the evolution of matter overdensities in Fourier space are as follows \citep{Tsujikawa_fR_dynamics2008PhRvD}: 
\begin{equation}
\begin{aligned}
& \ddot{\delta}_m+\left(2 H+\frac{\dot{F}}{2 F}\right) \dot{\delta}_m-\frac{\rho_m}{2 F} \delta_m \\
& =\frac{1}{2 F}\left[\left(-6 H^2+\frac{k^2}{a^2}\right) \delta F+3 H \delta \dot{F}+3 \delta \ddot{F}\right]\,, \\
& \delta \ddot{F}+3 H \delta \dot{F}+\left(\frac{k^2}{a^2}+\frac{F}{3 F_R}-4 H^2-2 \dot{H}\right) \delta F \quad=\frac{1}{3} \delta \rho_m+\dot{F} \dot{\delta}_m \,,
\end{aligned}\label{eq:deltaMG}
\end{equation}
with $H$ being the Hubble parameter, $k$ the comoving wavenumber of the perturbations, $a$ the scale factor, $\rho_m$ the matter density field, and $F \equiv \partial f /\partial R$. The solution to the system of Eqs. \eqref{eq:deltaMG} provides a detailed description of $\delta_m$, which includes most of the cosmological information since it is a direct result of the gravitational interaction of cosmic structures. To obtain insights into the underlying cosmic parameters, the density field is the primary source to be investigated using summary statistics. The Eqs. \eqref{eq:deltaMG} make evident the connection between the density field and the scalaron of MG. Consequently, any departure from GR would be measurable through the density field, either with the structure growth rate, or with its tracer distribution. A particular feature of the $f(R)$ models is the so-called chameleon mechanism. This mechanism reconciles the departures of GR with the bounds imposed by local tests of gravity. It gives the mass of the scalar field the ability to depend on the local matter density. More precisely, the signatures of MG can be detected in regions of lower matter density where the scalar field becomes lighter, leading to potentially observable effects that deviate from standard gravity~\citep{odintsov2023recent}.

The 3D matter power spectrum, denoted $P(k)$, is a primary statistical tool employed to extract cosmological insights from the density field. It characterises the overdensities as a function of scale and is estimated through the following average over Fourier space:
\begin{equation}
(2 \pi)^3 P(\boldsymbol{k}) \delta_{\mathrm{D}}^3\left(\boldsymbol{k}-\boldsymbol{k}^{\prime}\right)=\left\langle\delta(\boldsymbol{k}) \delta\left(\boldsymbol{k}^{\prime}\right)\right\rangle\,,
\end{equation}
where $\delta_{\mathrm{D}}^3$ is the 3D Dirac-delta function. This function contains all the information from the statistics of the density field in the linear regime and a decreasing fraction of the total information on smaller scales when the initial density fields follow Gaussian statistics. We used the \texttt{Pylians3}\footnote{\url{https://pylians3.readthedocs.io/en/master/index.html}} library to estimate the overdensity field as well as the power spectrum. 

%-------------------------------------------------------------------
\subsection{Modified-gravity simulations}
The simulations were created with the COmoving Lagrangian Acceleration (COLA) algorithm \citep{Tassev_COLA_2013JCAP, Koda_COLA2016}, which is based on a particle-mesh code that solves the equations of motion following the Lagrangian perturbation theory (LPT) trajectories of the particles. This algorithm speeds up the computation of the gravitational force using very few time steps and still obtains correct results on the largest scales. In particular, we used \texttt{MG-PICOLA}\footnote{The code can be found at \url{https://github.com/HAWinther/MG-PICOLA-PUBLIC}}~\citep{Winther-COLA_scaledependent2017}, a modified version of \texttt{L-PICOLA} \citep{2015A&C....12..109H} that has been extensively tested against full N-body simulations and that extends the gravity solvers to MG models, including the HS parametrisation. 
\begin{table}
      \caption[]{The summary of the setup of the MG simulations. Left: Cosmology parameters. Right: Setup parameters used for the \texttt{MG-PICOLA} code.}
         \label{tab:simsetup}
\begin{tabular}{ccccc}
\cline{1-2} \cline{4-5}
\multicolumn{2}{c}{\textbf{Cosmologies}}                                    &  & \multicolumn{2}{c}{\textbf{Simulation setup}}                           \\ \cline{1-2} \cline{4-5} 
\cellcolor[HTML]{EFEFEF}$\Omega_m$ & \cellcolor[HTML]{EFEFEF}{[}0.1, 0.5{]} &  & \cellcolor[HTML]{EFEFEF}Boxsize    & \cellcolor[HTML]{EFEFEF}$256$ \Mpch \\
$h$                                & {[}0.5, 0.9{]}                         &  & $N_p$                              & $128^3$                            \\
\cellcolor[HTML]{EFEFEF}$\sigma_8$ & \cellcolor[HTML]{EFEFEF}{[}0.6, 1.0{]} &  & \cellcolor[HTML]{EFEFEF}Grid force & \cellcolor[HTML]{EFEFEF}$128^3$    \\
$0.1\log_{10}|f_{R0}|$                 & {[}0.4, 0.6{]}                         &  & IC                                 & 2LPT $z_{ini}=49$                  \\
\cellcolor[HTML]{EFEFEF}$\Omega_b$ & \cellcolor[HTML]{EFEFEF}0.0489         &  & \cellcolor[HTML]{EFEFEF}Steps      & \cellcolor[HTML]{EFEFEF}100        \\
$n_s$                              & 0.9665                                 &  & $k_{Ny}$                           & 1.58                              \\ 
\cline{1-2} \cline{4-5}
\end{tabular}
\end{table}
\begin{figure}
\centering
\includegraphics[width=\linewidth]{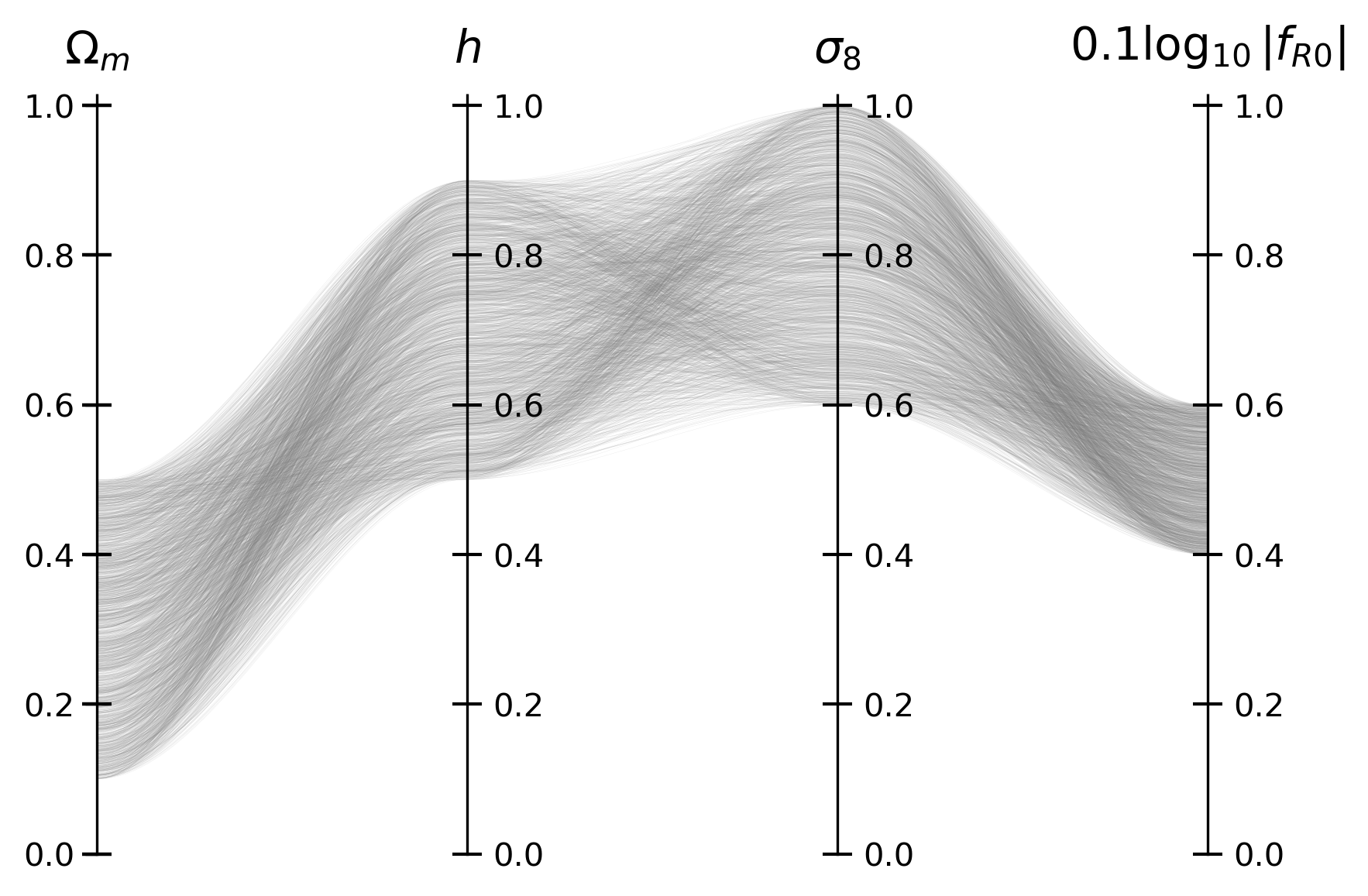}
\caption{Multidimensional parameter space variations. Each line represents the parameter values of a data point. The four parameters $\Omega_m,\, h,\,\sigma_8$, and  $0.1\log_{10}|f_{R0}|$ are visualised along separate axes.}
\label{fig:param_space}%
\end{figure}
\begin{figure}
    \centering
    \includegraphics[width=\linewidth]{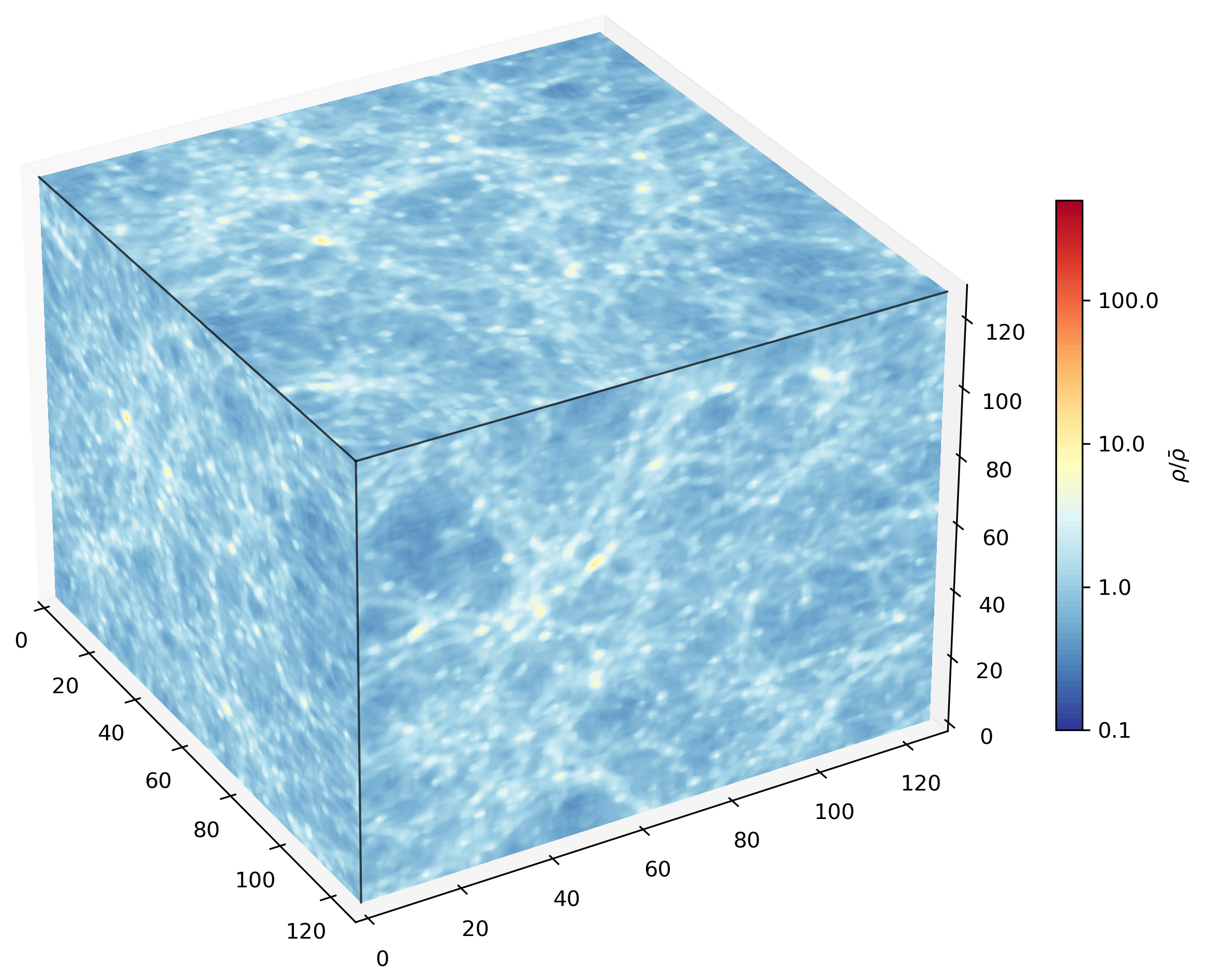}
    \caption{Projected overdensity field at redshift $z=0$ derived from an arbitrarily chosen simulation within the ensemble of 2500 MG simulations. The normalised density field was calculated using a CIC mass-assignment scheme.}
    \label{fig:cube_voxels128}%
\end{figure}
\begin{figure*}
   \resizebox{\hsize}{!}
    {\includegraphics[width=\textwidth]{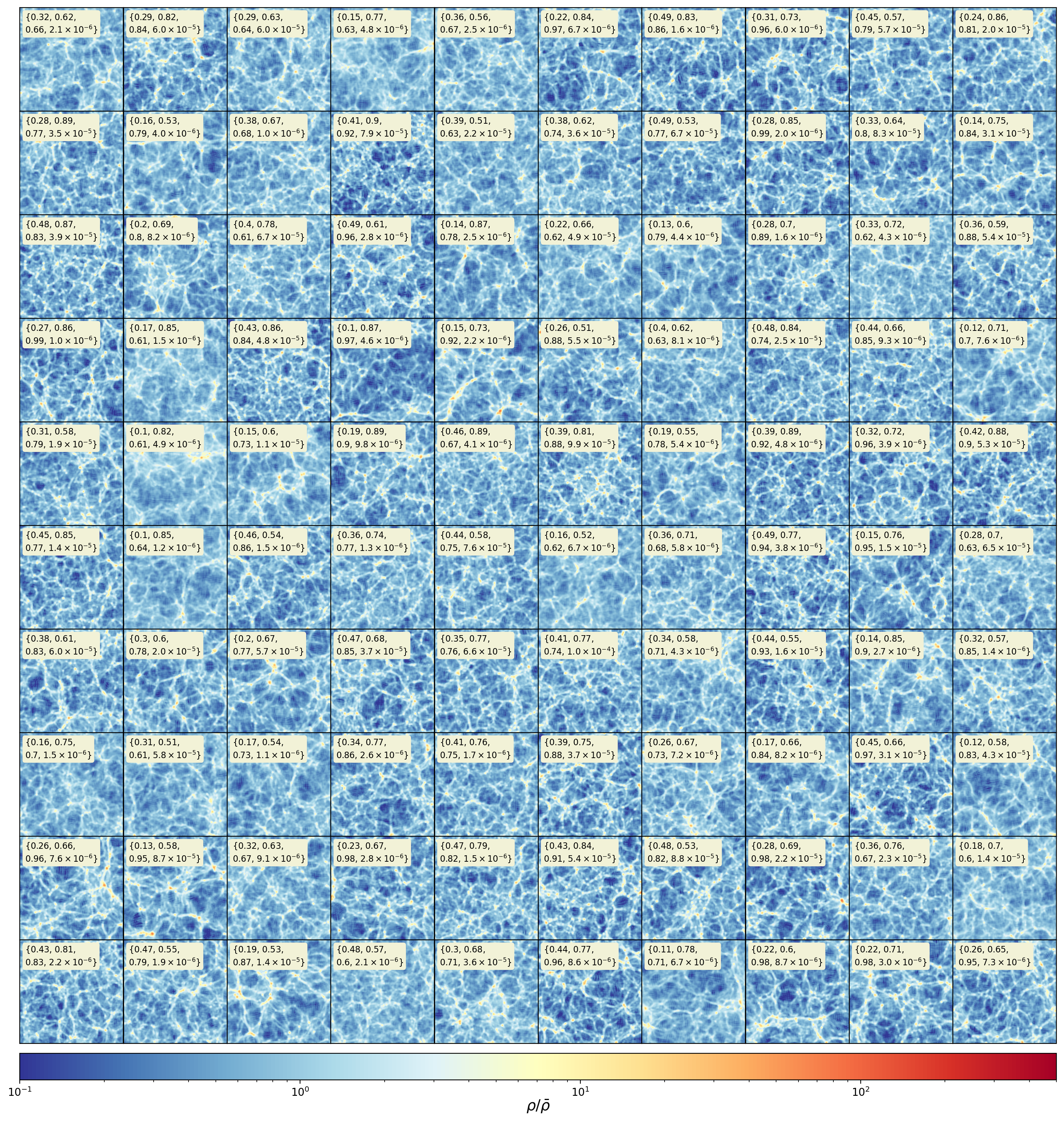}}
    \caption{Projected density field of dark matter in a region of $256\times256\times20$ ($h^{-1}Mpc$)${}^3$ from 100 out of 2500 arbitrarily chosen simulations of MG. The snapshots are taken at $z=0$, and the legend displays the set of cosmological parameters of \{$\Omega_m$, $h$, $\sigma_8$, $f_{R_0}$\}. The cuts in the density field highlight the broad coverage of the parameter space of the MG simulations. Different features can be observed by naked eye, such as variations in the filament structure of the cosmic web.}
    \label{fig:projected_field_100boxes}
\end{figure*}
We ran a set of 2500 MG simulations for which we varied the four cosmological parameters $\Theta=\{\Omega_m,\, h,\,\sigma_8,\, f_{R0}\}$, where $h$ is the reduced Hubble parameter, $\sigma_8$ is the r.m.s. density fluctuation within a top-hat sphere with a radius of $8$ \Mpch\, and $f_{R0}$ is the amplitude of the MG function in the HS model. The remaining cosmological parameters were set to $\Omega_b = 0.048206$ and $n_s = 0.96$, which correspond to the values reported by \citet{Planck_parameters2020}. The parameter space was sampled with random numbers uniformly distributed within the specified ranges for each parameter (see \cref{tab:simsetup}). Since the typical values of the MG parameter go as powers of ten, $|f_{R0}|\sim10^n$ with $n\in[-4,-6]$, we chose to sample a fraction of its logarithm in order to cover the range of powers equally, that is, $\widetilde{f}_{R0}=0.1\log_{10}|f_{R0}|$. Figure \ref{fig:param_space} illustrates the parameter space variations of the 2500 MG cosmologies, each of which is represented by a grey line. The values of the cosmological parameters, $\Theta$, are distributed along the different vertical axes.
Each simulation follows the dynamics of $128^3$ particles in a small box with a comoving side-length of $256$ \Mpch\ using 100 time steps from an initial redshift $z_i=49$ up to redshift $z=0$. This simulation resolution allows us to reasonably investigate the impacts of MG on large scales, in particular for the $f_{R0}$ values considered in this work. However, it is not as effective at very small scales, where a higher resolution is required. MG solvers have undergone extensive testing using low-resolution simulations (see e.g. \cite{2013MNRAS.436..348P, 2012JCAP...01..051L, 2022JCAP...01..048H}). These tests show the enhancement of the power spectrum in simulations of $256$ \Mpch, where MG effects begin to be seen. Our setup of the MG simulations is summarised in \cref{tab:simsetup}. In a recent research, a similar setup was employed with a light-weight deterministic CNN to estimate a subset of parameters of a flat \lcdm\ cosmology \citep{2020SCPMA..6310412P}, but we chose a time-step that is larger by a factor of 2.5. The initial conditions for the MG simulations were created with \texttt{2LPTic} \citep{2006MNRAS.373..369C,2012ascl.soft01005C} based on a \lcdm\ template at $z_i$. Moreover, a distinct random seed was assigned to each simulation to generate varying distributions of large-scale power. This approach allowed our neural network to effectively capture the inherent cosmic variance.

\begin{figure}[h!]
    \centering
    \includegraphics[width=\linewidth]{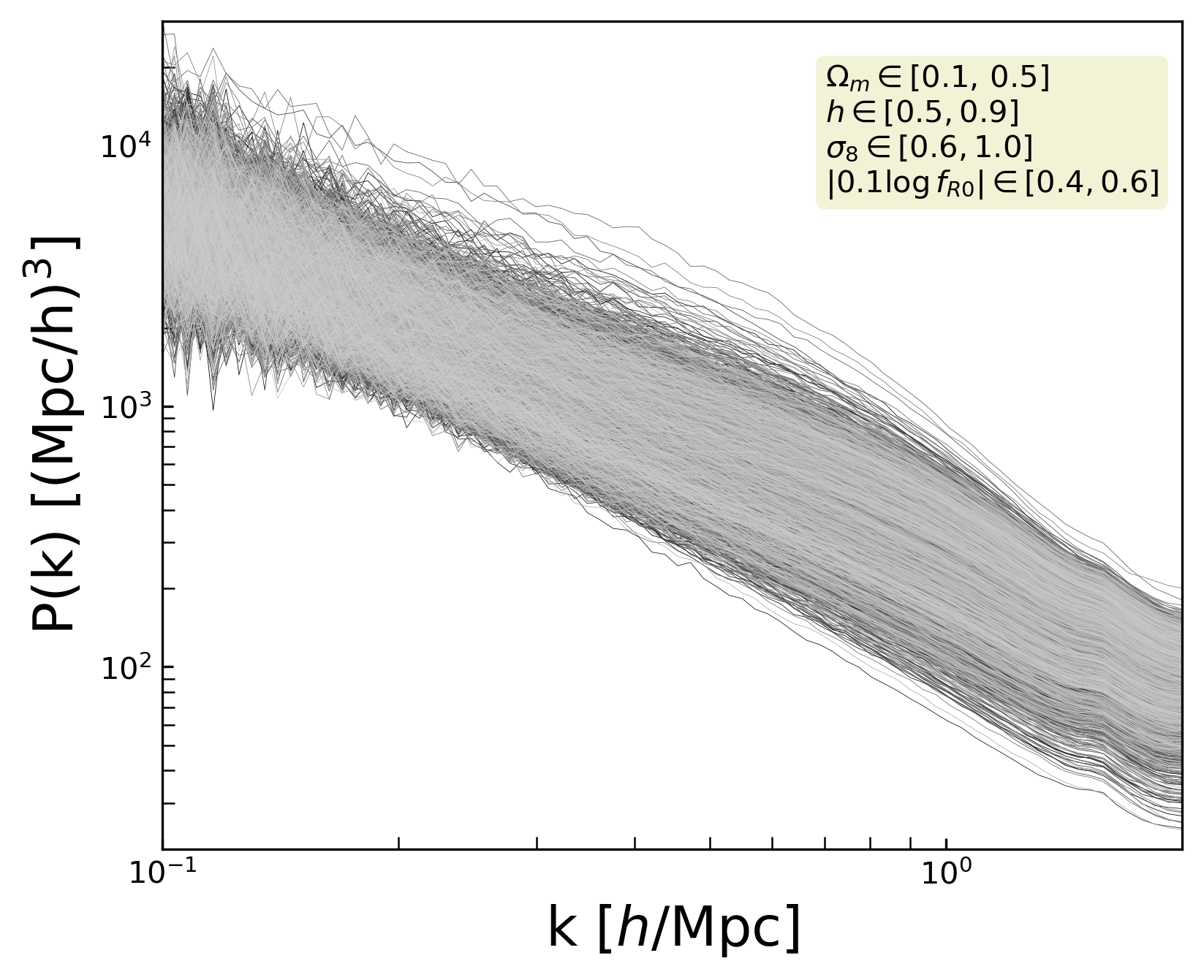}
    \caption{Matter power spectrum at $z=0$ of the MG simulation suit. The variations in the spectrum correspond to changes in each of the four parameters that were varied, $\Omega_m$, $h$, $\sigma_8$ and $|0.1\log f_{R0}|$. The corresponding range of each of parameter is shown in the label.}
    \label{fig:pks_sim}%
    \end{figure}

We calculated the overdensity field, $\delta_{\rm m}$, for each snapshot at redshift $z=0$ employing the cloud-in-cell (CIC) mass-assignment scheme \citep{1981csup.book.....H} on a regular grid consisting of $N^3=128^3$ voxels. The training set comprised 80\% of the data, which corresponds to 2000 boxes containing the overdensity fields, while the remaining 20\% of the data were used for testing. Fig. \ref{fig:cube_voxels128} displays the 3D overdensity field plus the unity, $\delta_m+1=\rho_m/\bar{\rho}_m$, projected along each plane of the box. The displayed data correspond to an arbitrarily chosen combination of parameters within the MG simulation suite at redshift $z=0$. Similarly, Fig. \ref{fig:projected_field_100boxes} displays the 2D density field of dark matter in a region of $256\times256\times20$ ($h^{-1}Mpc$)${}^3$ from 100 out of 2500 arbitrarily chosen simulations of MG. The cosmological parameter combination is indicated by the labels. The cuts in the density field provide a visual means to discern distinct features of the cosmic web that are observable with the naked eye. These features include variations in the filament structure of the cosmic web, which become evident in the zones of under- and over-densities. Additionally, we output the matter power spectrum of all realisations by directly computing the modulus of each Fourier mode from the particle distribution, $|\delta_m(k)|^2$. The \cref{fig:pks_sim} shows the different matter power spectra for the entire MG simulation suit. The variations in the shape of the spectrum correspond to the joint effect of cosmological parameters that were varied as shown in the label. We considered the effective range of the power spectrum up to the Nyquist frequency, $k_{Ny}$, which in our simulations corresponds to $k \approx 1.58$ \MpchInv. The full datasets used in this paper, 3D overdensity fields as well as power spectra, are publicly available in~\href{https://zenodo.org/}{\texttt{Zenodo}}\footnote{\href{https://doi.org/10.5281/zenodo.10555349}{https://doi.org/10.5281/zenodo.10555349}}.
  
%-------------------------------------------------------------
\section{Bayesian neural networks}\label{secbbn}
The primary goal of BNNs is to estimate the posterior distribution $p(w|\calD)$, which represents the probability distribution of the weights $w$ of the network given the observed data $\calD=(X, Y)$~\citep{Abdar_2021,Gal2016Uncertainty,NIPS2011_7eb3c8be}.  The posterior distribution, denoted as $p(w|\calD)$, can be derived using Bayes' law: $p(w|\calD)\sim p(\calD|w)p(w)$. This expression involves a likelihood function, $p(\calD | w)$, which represents the probability of the observed data $\calD$ given the weights $w$, as well as a prior distribution on the weights, denoted as $p(w)$. After the computation of the posterior, the probability distribution of a new test example $x^*$ can be determined by
\begin{equation}
  p( y^* |x^*, \calD) = \int_w p(y^*|x^*,w)p(w|\calD)dw\,,
\end{equation}
where $p(y^*|x^*, w)$  is the predictive distribution corresponding to the set of weights. In the context of neural networks, it is important to note that the direct computation of the posterior is not feasible~\citep{Gal2016Uncertainty}. To circumvent this limitation, variational inference (VI) techniques approximating the posterior distribution were introduced \citep{NIPS2011_7eb3c8be}. VI considers a family of simple distributions, denoted as $q(w | \theta)$, which is characterised by a parameter $\theta$. The objective of VI is to identify a distribution $q(w | \theta^*)$ that minimises the Kullback-Leibler divergence between $q(w|\theta)$ and $p(w|\calD)$, where $\theta^*$ represents the optimal parameter values, and $KL[\cdot\|\cdot]$ is the Kullback-Leibler divergence. This optimisation  is equivalent to maximising the evidence lower bound (ELBO)~\citep{Gal2016Uncertainty}, 
\begin{equation}\label{elbo}
  \ELBO(\theta) = \erw_{q(w|\theta)}\big[\log p(Y|X,w)\big] - \KL\big[q(w|\theta)\big\|p(w)\big]\,,
\end{equation}
where $\erw_{q(w|\theta)}[\log p(Y|X,w)]$ is the expected log-likelihood with respect to the variational posterior and $\KL[q(w|\theta)||p(w)]$ is the divergence of the variational posterior from the prior.
Eq.~\eqref{elbo}  shows that the Kullback-Leibler (KL) divergence serves as regulariser, compelling the variational posterior to shift towards the modes of the prior.
A frequently employed option for the variational posterior entails using a product of independent Gaussian distributions, specifically, mean-field Gaussian distributions. Each parameter $w$ is associated with its own distribution~\citep{Abdar_2021},
\begin{equation}\label{meanfield}
    q(w|\theta)=\prod_{ij} \calN(w ;\mu_{ij},\sigma^2_{ij})  \,,
\end{equation}
where $i$ and $j$ are the indices of the neurons from the previous  and current layers, respectively. Applying the reparametrisation trick,  we obtained $w_{ij}=\mu_{ij}+\sigma_{ij}*\epsilon_{ij}$, where $\epsilon_{ij}$ was drawn from the normal distribution. Moreover, when the prior is a composition of independent Gaussian distributions, the KL-divergence between the prior and the variational posterior can be calculated analytically. This characteristic enhances the computing efficiency of this approach. 
\subsection{Multiplicative normalising flows}\label{secbbn4}
The Gaussian mean-field distributions described in  Eq.~\eqref{meanfield} are the most commonly used family for the variational posterior in BNNs. Unfortunately, this distribution lacks the capacity to adequately represent the intricate nature of the true posterior. Hence, it is anticipated that enhancing the complexity of the variational posterior will yield substantial improvements in performance. This is attributed to the capability of sampling from a more reliable distribution, which closely approximates the true posterior distribution. The process of improving the variational posterior demands efficient computational methods while ensuring its numerical feasibility. Multiplicative normalising flows (MNFs) have been proposed to efficiently adapt the posterior distributions by using auxiliary random variables and the normalising flows~\citep{10.5555/3305890.3305910}. Mixture normalising flows (MNFs) suggest that the variational posterior can be mathematically represented as an infinite mixture  of distributions,
\begin{equation}\label{mixture}
    q(w|\theta)= \int q(w|z,\theta)q(z|\theta)dz \,,
\end{equation}
with $\theta$  the learnable posterior parameter, and $z\sim q(z|\theta)\equiv q(z)$\footnote{For the sake of clarity in notation, the parameter $\theta$ will no longer be considered in the subsequent discussion.} the vector with the same dimension as the input layer,  which plays the role of an auxiliary latent variable. Moreover, allowing for local reparametrisations, the variational posterior for fully connected layers becomes
\begin{equation}\label{mnfpos}
   w \sim q(w|z)=\prod_{ij}\calN(w;z_i\mu_{ij},\sigma^2_{ij})\,.
\end{equation}
Here, we can increase the flexibility of the variational posterior by enhancing the complexity of $q(z)$. This can be done using normalising flows since the dimensionality of $z$ is much lower than the weights. Starting from samples $z_0\sim q(z_0)$ from fully factorised Gaussians (see Eq.~\eqref{meanfield}), a rich distribution $q(z_K)$ can be obtained by applying  successively invertible $f_k$-transformations, 
\begin{equation}
\label{nfz0}
z_K=\text{NF}(z_0)=f_K \circ\cdots\circ f_1(z_0)\,,
\end{equation}
\begin{equation}
\label{nfz01}
\log q(z_K)=\log q(z_0) - \sum_{k=1}^K\log \left|\det\frac{\partial f_k}{\partial z_{k-1}} \right|\,.
\end{equation}
To handle the intractability of the posterior, \citet{10.5555/3305890.3305910} suggested to use again Bayes law $q(z_K)q(w|z_K)=q(w)q(z_K|w)$ and introduce a new auxiliary distribution $r(z_K | w,\phi)$ parametrised by $\phi$, with the purpose of approximating the posterior distribution of the original variational parameters $q(z_K|w)$  to further lower the bound of the KL divergence term. Accordingly, the  KL-divergence term  can be rewritten as follows:
\begin{align}\label{velbo}
 - \KL\big[q(w)\big\|p(w)\big] \geq \erw_{q(w,z_K)}\biggl[ - \KL\big[q(w|z_K)\big\|p(w)\big]\biggr. \nonumber \\ \biggl. + \log q(z_K)  +\log r(z_K|w,\phi)     \biggr]\,.
\end{align}
The first term can be analytically computed since it will be the KL-divergence between two Gaussian distributions, while the second term is computed via the normalising flow generated by $f_K$ (see Eq.~\eqref{nfz01}). Furthermore, the auxiliary posterior term is parametrised by inverse normalising flows as follows~\citep[][]{1806.02315}:
\begin{equation}\label{nfz1}
z_0=\text{NF}^{-1}(z_K)=g^{-1}_1 \circ\cdots\circ g^{-1}_K(z_K)\,, 
\end{equation}
and
\begin{equation}\label{nfz12}
\log r(z_K|w,\phi)=\log r(z_0|w,\phi) + \sum_{k=1}^K\log \left|\det\frac{\partial g^{-1}_k}{\partial z_{k}} \right|\,,
\end{equation}
where $g^{-1}_K$ can be parametrised as another normalising flow. A flexible parametrisation of the auxiliary posterior can be given by
\begin{equation}
    z_0\sim  r(z_K|w,\phi)=\prod_i \calN(z_0;\tilde{\mu}_i(w,\phi),\tilde{\sigma}^2_i(w,\phi))\,,
\end{equation}
where  the parameterisation of  the mean  $\tilde{\mu}$, and the variance $\tilde{\sigma}^2$  is carried out by the masked RealNVP~\citep{dinh2017density} as the choice of normalising flows. \\
\subsection{Multiplicative normalising flows in a voxel-grid representation}\label{secbbn43d}
In this section, we present our first result, where we have generalised  Eq.~\eqref{mnfpos} to 3D convolutional layers where cosmological simulated data are structured. To this end, we started with the extension of sampling from the variational posterior as
\begin{equation}\label{mnfpos3d}
   w \sim q(w|z)=\prod_{i}^{D_d}\prod_{j}^{D_h}\prod_{k}^{D_w}\prod_{l}^{D_f}\calN(w;z_l\mu_{ijkl},\sigma^2_{ijkl})\,,
\end{equation}
\begin{algorithm}[htb]
\caption{Forward propagation for each convolutional 3D layer. $M_w$, $\Sigma_w$ are the means and variances of each layer, $H$ is the input layer, and NF(·) is the masked
RealNVP normalising flow applied over samples initially drawn from a Gaussian distribution $q$. $D_f$ is the number of filters for each kernel. $\odot$ corresponds to element-wise multiplication.} 
\label{alg:conv_bnn3d}
\begin{algorithmic}
\State $\text{Input: feature vector of the previous layer (minibatch)}$
\State $H \gets \text{Input conv3D-layer (minibatch)}$
    \State $\*z_0 \sim q(\*z_0)$
    \State $\*z_{T_f} = \text{NF}(\*z_0)$
    \State $\*M_h = \*H * (\*M_w \odot \text{reshape}(\*z_{T_f}, [1,1, 1, D_f]))$
    \State $\*V_h = \*H^2 * \!\Sigma_w$
    \State $\*E \sim \mathcal{N}(0, 1)$
    \State return $\*M_h + \sqrt{\*V_h} \odot \*E$
    \State$\text{Output: sample of feature vector according to Eq.~\ref{mnfpos3d}}$
\end{algorithmic}
\end{algorithm}
where $D_h$, $D_w$, and $D_d$ are the three spatial dimensions of the boxes,  and $D_f$ is the number of filters for each kernel. The objective is to address the challenge of enhancing the adaptability of the approximate posterior distribution for the weight coming from a 3D convolutional layer.  Algorithm \ref{alg:conv_bnn3d} outlines the procedure to forward propagation for each 3D convolutional layer. Similar to the fully connected case, the auxiliary parameter only affects the mean with the purpose of avoiding large variance, and  we kept a linear mapping to parameterise the inverse
normalising flows instead of applying $\tanh$ feature transformations. 

%-------------------------------------------------------------------
\section{The Bayesian architecture setup}\label{sec:architectures1}
\begin{figure}
   \centering
    \includegraphics[width=\linewidth]{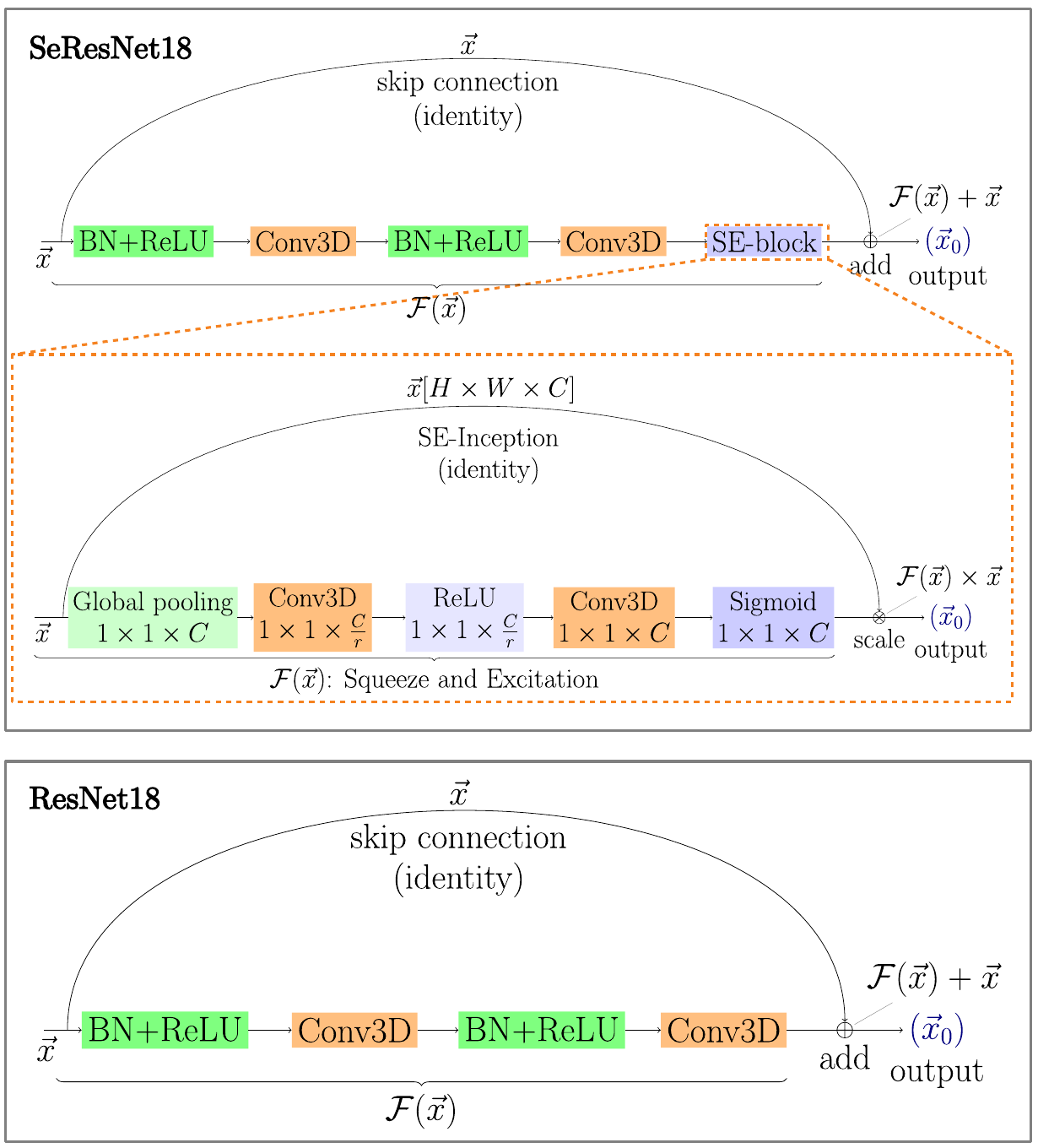}
    \caption{Resblock schema depending on the architecture used. Top: Resblock when SeResNet18 is employed. The dashed orange rectangle shows the skip SE-connection schema used in the SeResNet18 resblock. Bottom:  Resblock when ResNet is employed.}
    \label{fig:skip-connections}
   \end{figure}

We examined four distinct architectures of BNNs, as outlined in Section~\ref{secbbn}. Two of these architectures include Bayesian layers located only on top of the network, the so-called Bayesian last layer (denoted as \textbf{BLL}), while the remainder have Bayesian layers at all their levels (\textbf{FullB}). The pipelines used in our study were developed using TensorFlow v:2.12\footnote{\url{https://www.tensorflow.org/}} and TensorFlow-probability v:0.19\footnote{\url{https://www.tensorflow.org/probability}}~\citep{tensorflow2015-whitepaper}. The architecture used for all networks has  ResNet-18 as the backbone, which is depicted in a schematic manner in Table~\ref{tab:BNN_}. The input layer receives simu\-lated boxes of $(128,128,128,1)$ with normalised voxels in the range 0 and 1.
\begin{table}
\caption{Configuration of the (Se)-ResNet backbone  used for all experiments presented in this paper.}\label{tab:BNN_}
\tiny
\centering
\begin{tabular}{ccc}
\hline
 \multicolumn{3}{c}{ \textbf{(Se)-ResNet-18 backbone} } \\ [0.1cm]
\cellcolor[HTML]{EFEFEF}\textbf{Layer Name} & \cellcolor[HTML]{EFEFEF}\textbf{Input Shape} & \cellcolor[HTML]{EFEFEF}\textbf{Output Shape} \\ [0.1cm] \hline
Batch Norm & ($N_\text{batch}$, 128,128,128,1) & ($N_\text{batch}$, 128,28,128,1) \\ [0.1cm]
\cellcolor[HTML]{EFEFEF}3D Convolutional & \cellcolor[HTML]{EFEFEF}($N_\text{batch}$, 128,128,128,1) & \cellcolor[HTML]{EFEFEF}($N_\text{batch}$, 64,64,64,16) \\  [0.1cm]
Batch Norm+ReLU & ($N_\text{batch}$, 64,64,64,16) & ($N_\text{batch}$, 64,64,64,16) \\  [0.1cm]
\cellcolor[HTML]{EFEFEF}Max Pooling 3D & \cellcolor[HTML]{EFEFEF}($N_\text{batch}$, 64,64,64,16) & \cellcolor[HTML]{EFEFEF}($N_\text{batch}$, 32,32,32,16) \\  [0.1cm]
Batch Norm+ReLU & ($N_\text{batch}$, 32,32,32,16 ) & ($N_\text{batch}$, 32,32,32,16 ) \\ [0.1cm]
\cellcolor[HTML]{EFEFEF}Resblock 1& \cellcolor[HTML]{EFEFEF}$\begin{bmatrix}
(N_\text{batch},  32,32,32,16) \\
(N_\text{batch},  16,16,16,32) \end{bmatrix}$  & \cellcolor[HTML]{EFEFEF}($N_\text{batch}$,  16,16,16,32)\\ [0.5cm]
Batch Norm+ReLU & ($N_\text{batch}$,  16,16,16,32) & ($N_\text{batch}$,  16,16,16,32 ) \\  [0.3cm]
\cellcolor[HTML]{EFEFEF}Resblock 2& \cellcolor[HTML]{EFEFEF}$\begin{bmatrix}
(N_\text{batch},  16,16,16,32) \\
(N_\text{batch},  8,8,8,64) \end{bmatrix}$  & \cellcolor[HTML]{EFEFEF}($N_\text{batch}$, 8,8,8,64)\\ [0.5cm]
Batch Norm+ReLU & ($N_\text{batch}$, 8,8,8,64) & ($N_\text{batch}$, 8,8,8,64 ) \\ [0.3cm] 
\cellcolor[HTML]{EFEFEF}Resblock 3& \cellcolor[HTML]{EFEFEF}$\begin{bmatrix}
(N_\text{batch},  8,8,8,64 ) \\
(N_\text{batch},  4,4,4,128) \end{bmatrix}$  & \cellcolor[HTML]{EFEFEF}($N_\text{batch}$, 4,4,4,128)\\[0.5cm]
Batch Norm+ReLU & ($N_\text{batch}$, 4,4,4,128) & ($N_\text{batch}$, 4,4,4,128) \\  [0.3cm]
\cellcolor[HTML]{EFEFEF}Resblock 4& \cellcolor[HTML]{EFEFEF}$\begin{bmatrix}
(N_\text{batch},   4,4,4,128) \\
(N_\text{batch},  2,2,2,256) \end{bmatrix}$  & \cellcolor[HTML]{EFEFEF}($N_\text{batch}$,  2,2,2,256)\\ [0.5cm]
Batch Norm+ReLU & ($N_\text{batch}$, 2,2,2,256 ) & ($N_\text{batch}$,   2,2,2,256) \\  [0.1cm]
\cellcolor[HTML]{EFEFEF}Global Avg Pooling & \cellcolor[HTML]{EFEFEF}($N_\text{batch}$, 2,2,2,256) & \cellcolor[HTML]{EFEFEF}($N_\text{batch}$, 256) \\ 
\hline
\end{tabular}
\end{table}
The Resblock nature depends on whether we build a ResNet or an SeResNet topology.  The latter is a variant of ResNet that employs squeeze-and-excitation blocks that adaptively re-calibrate channel-wise feature responses by explicitly modelling inter-dependences between channels~\citep{hu2019squeezeandexcitation}. Fig.~\ref{fig:skip-connections}  depicts each   Resblock and how the skip connections are defined. These architectures were designed using the \texttt{GIT} repository classification-models-3D\footnote{\url{https://github.com/ZFTurbo/classification_models_3D}}. 
ResNet18 contains 2510149 trainable parameters while SeResNet has 3069270, but these numbers are duplicates in a fully Bayesian scheme because two parameters need to be optimised (the mean and standard deviation) for each network parameter.
In this study, 50 layers were employed for the masked RealNVP normalising flow. The development of these convolutional layers was made using the repositories TF-MNF-VBNN\footnote{\url{https://github.com/janosh/tf-mnf/tree/0ed492bd8faf0bdc37a56f87adf2d8ca425eec5b}} and MNF-VBNN\footnote{\url{https://github.com/AMLab-Amsterdam/MNF_VBNN}} as well ~\citep{10.5555/3305890.3305910}. Finally, all networks end with a  multivariate Gaussian distribution layer, consisting of 14 trainable parameters. These parameters include four means, denoted as $\mathbf{\mu}$, which correspond to the cosmological parameters, as well as ten elements representing the covariance matrix $\Sigma$. TensorFlow probability has a built-in version of the multivariate normal distribution, MultivariateNormalTriL, that is parametrised in terms of the Cholesky factor of the covariance matrix. This decomposition guarantees the positiveness of the covariance matrix. In addition, we  included a reparametrisation led by ReLU activation to force positive values in the mean of the distribution. The loss function to be optimised  is given by the ELBO, ~\cref{elbo}, where the second term is associated with the negative log-likelihood (NLL), 
 \begin{equation}\label{nll}
     -\text{NLL}\sim \frac{1}{2}\log|\Sigma|+\frac{1}{2}(\mathbf{y}-\mathbf{\mu})^\top
     \left(\Sigma\right)^{-1} (\mathbf{y}-\mathbf{\mu})\,,
 \end{equation}
averaged over the mini-batch. The optimiser used was an Adam optimiser with first- and second-moment exponential decay rates of 0.9 and 0.999, respectively~\citep[][]{https://doi.org/10.48550/arxiv.1412.6980}. The learning rate started from $5\times 10^{-4}$ and was reduced by a factor of $0.9$ when no improvement was observed after 8 epochs. Furthermore, we applied a warm-up period for which the model progressively turned on the KL term in Eq.~\eqref{elbo}. This was achieved by introducing a $\beta$ variable in the ELBO, that is, $\beta\cdot\KL\big[q(w|\theta)\big\|p(w)\big]$, so that this parameter started being equal to 0 and grew linearly to 1 during 12 epochs~\citep[][]{3157382.3157516}. BNNs were trained with a batch size of 8, and early-stopping callback was presented to avoid overfitting. Finally, hyperparameter tuning was applied to the filters (the initial filters were 64, 32, and16, after which they increased as $2^n$, where $n$ stands for the number of layers) in the convolutional layers for ResNet (18,32,50), SeResNet (18,32,50), VGG(16,19), and MobileNet.  The infrastructure set in place by the Google Cloud Platform\footnote{\url{https://cloud.google.com/}} (GCP) uses an nvidia-tesla-t4 of  16 GB GDDR6 in an N1 machine-series shared-core.   
\subsection{Quantifying the performance}\label{subsec:performance}
The metrics employed for determining the network performance were the mean square error (MSE),  ELBO, and the coefficient of the determination $r^2$.  Moreover, we quantified the quality of the uncertainty estimates through reliability metrics. Following~\citet{laves2020wellcalibrated} and \citet{Guo:2017:CMN:3305381.3305518}, we defined a perfect calibration of the regression uncertainty as
\begin{equation}
    \mathbb{E}_{\hat{\sigma}^{2}} \left[ \text{abs}\big[\big( ||\bm{y}-\bm{\mu}||^2 \, \big| \, \hat{\sigma}^{2} = \alpha^{2} \big) - \alpha^{2} \big] \right] \quad \forall \left\{ \alpha^{2} \in \mathbb{R} \, \big| \, \alpha^{2} \geq 0 \right\}\,,
    \label{eq:uce}
\end{equation}
where $\text{abs}[.]$ is the absolute value function. The predicted uncertainty  $ \hat{\sigma}^{2} $  was partitioned into $ K $ bins with equal width in this way, and the variance per bin is defined as
\begin{equation}
    \mathrm{var}(B_{k}) := \frac{1}{\big| B_{k} \big|} \sum_{i \in B_{m}} {\frac{1}{N} \sum_{n=1}^{N} \left( \bm{\mu}_{i,n} - \bm{y}_{i} \right)^{2} } \,,
\end{equation}
with $ N $ stochastic forward passes. In addition,  the uncertainty per bin is defined as
\begin{equation}
    \mathrm{uncert}(B_{k}) := \frac{1}{\vert B_{k} \vert} \sum_{i \in B_{k}} \hat{\sigma}_{i}^{2}\,,
\end{equation}
which allowed us to compute the expected uncertainty calibration error (UCE) in order to quantify  the miscalibration, 
\begin{equation}\label{uce}
    \mathrm{UCE} := {\sum_{k=1}^{K}} \frac{\vert B_{k} \vert}{m} \big| {\mathrm{var}}(B_{k}) - \mathrm{uncert}(B_{k}) \big| \,,
\end{equation}
with the number of inputs $m$ and set of indices $ B_{k} $ of inputs, for which the uncertainty falls into bin $ k $. 

%-------------------------------------------------------------------
\section{Results}
In this section, we present the results of several experiments we developed to quantify the performance of Bayesian deep-learning neural networks for constraining the cosmological parameters in MG scenarios. 

\subsection{Parameter estimation from the overdensity field under a voxel-grid representation}
Using the configuration described in Sec.~\ref{sec:architectures1}, we designed four experiments inspired by two successful deep-learning architectures, ResNet18 and Se-ResNet18. The former is a residual network commonly known due to its efficiency in several computer vision tasks, while the latter was chosen because of its ability to improve the interdependences between the channels of convolutional feature layers~\citep{hu2019squeezeandexcitation}. Furthermore,  the modification of the models was also based on the insertion of a set of Bayesian layers at either the top of the model (BLL) or in the entire architecture (FullB). The motivation for exploring both possibilities comes from the fact that intuitively, adding a Bayesian layer at the end of the network (BLL) can be viewed as Bayesian linear regression with a learnable projected feature space, allowing for a successful balance between scalability and the degree of  model agnosticism~\citep{fiedler2023improved,pmlr-v130-watson21a}. Conversely, although fully Bayesian networks (FullB) would demand high computational resources, it has been reported that their Bayesian hidden layers are susceptible to out-of-distribution (OOD) examples that might improve the predictive uncertainty estimates~\citep{henning2021bayesian}. 
\begin{table*}[h!]
\caption{Metrics for the test set for all BNNs architectures. Top: SeResNet18. Bottom: ResNet18. High UCE values indicate miscalibration. The bold text is the minimum (maximum) value, $\downarrow$ ($\uparrow$) as indicated in the metric name, among the different parameters.}
\label{tab:dfield_ReSeResNet18a}
\resizebox{\linewidth}{!}{%
\begin{tabular}{l cccccc cccccc}
\hline
\multirow{2}{*}{\textbf{Metrics}} & \multicolumn{6}{c|}{\textbf{FullB-SeResNet18}} & \multicolumn{6}{c}{\textbf{BLL-SeResNet18}} \\

         & \cellcolor[HTML]{EFEFEF}$\Omega_m$ & \cellcolor[HTML]{EFEFEF}$h$ & \cellcolor[HTML]{EFEFEF}$\sigma_8$ & \cellcolor[HTML]{EFEFEF}$0.1\log_{10}|f_{R0}|$ & \cellcolor[HTML]{EFEFEF}$\Omega_mh^2$ & \multicolumn{1}{c|}{\cellcolor[HTML]{EFEFEF}$\sigma_8\Omega_m^{0.25}$} & \cellcolor[HTML]{EFEFEF}$\Omega_m$ & \cellcolor[HTML]{EFEFEF}$h$ & \cellcolor[HTML]{EFEFEF}$\sigma_8$ & \cellcolor[HTML]{EFEFEF}$0.1\log_{10}|f_{R0}|$ & \cellcolor[HTML]{EFEFEF}$\Omega_mh^2$ & \cellcolor[HTML]{EFEFEF}$\sigma_8\Omega_m^{0.25}$ \\
\hline
MSE $\downarrow$ & 0.001 & 0.01 & 0.0007 & \multicolumn{1}{c}{0.003} & 0.0009 & \multicolumn{1}{c|}{0.0008} & 0.003 & 0.013 & 0.0012 & \multicolumn{1}{c}{0.0035} & 0.0009 & 0.0013 \\
\cellcolor[HTML]{EFEFEF}$r^2\,\uparrow$ & \cellcolor[HTML]{EFEFEF}0.86 & \cellcolor[HTML]{EFEFEF}0.15 & \cellcolor[HTML]{EFEFEF}0.94 & \multicolumn{1}{c}{\cellcolor[HTML]{EFEFEF}0.04} & \cellcolor[HTML]{EFEFEF}0.85 & \multicolumn{1}{c|}{\cellcolor[HTML]{EFEFEF}0.93} & \cellcolor[HTML]{EFEFEF}0.80 & \cellcolor[HTML]{EFEFEF}0.03 & \cellcolor[HTML]{EFEFEF}0.90 & \multicolumn{1}{c}{\cellcolor[HTML]{EFEFEF}0.008} & \cellcolor[HTML]{EFEFEF}0.85 & \cellcolor[HTML]{EFEFEF}0.89 \\
UCE $\downarrow$ & 0.07 & \textbf{0.07} & \textbf{0.08} & \multicolumn{1}{c}{0.02} & 0.08 & \multicolumn{1}{c|}{\textbf{0.12}} & 0.03 & 0.3 & \textbf{0.08} & \multicolumn{1}{c}{0.05} & \textbf{0.022} & \textbf{0.15} \\ \cline{2-13} 
\cellcolor[HTML]{EFEFEF}AV-MSE $\downarrow$ & \multicolumn{6}{c|}{\cellcolor[HTML]{EFEFEF}\textbf{0.0043}} & \multicolumn{6}{c}{\cellcolor[HTML]{EFEFEF}0.0051} \\
NLL $\downarrow$ & \multicolumn{6}{c|}{-99.21} & \multicolumn{6}{c}{-3.38} \\\hline
\cellcolor[HTML]{EFEFEF}Inf.Time [ms] & \multicolumn{6}{c|}{\cellcolor[HTML]{EFEFEF}397} & \multicolumn{6}{c}{\cellcolor[HTML]{EFEFEF}290}  \\\hline

&&&&&&&&&&&&\\ 

\hline  
\multirow{2}{*}{\textbf{Metrics}} & \multicolumn{6}{c|}{\textbf{FullB-ResNet18}} & \multicolumn{6}{c}{\textbf{BLL-ResNet18}} \\

         & \cellcolor[HTML]{EFEFEF}$\Omega_m$ & \cellcolor[HTML]{EFEFEF}$h$ & \cellcolor[HTML]{EFEFEF}$\sigma_8$ & \cellcolor[HTML]{EFEFEF}$0.1\log_{10}|f_{R0}|$ & \cellcolor[HTML]{EFEFEF}$\Omega_mh^2$ & \multicolumn{1}{c|}{\cellcolor[HTML]{EFEFEF}$\sigma_8\Omega_m^{0.25}$} & \cellcolor[HTML]{EFEFEF}$\Omega_m$ & \cellcolor[HTML]{EFEFEF}$h$ & \cellcolor[HTML]{EFEFEF}$\sigma_8$ & \cellcolor[HTML]{EFEFEF}$0.1\log_{10}|f_{R0}|$ & \cellcolor[HTML]{EFEFEF}$\Omega_mh^2$ & \cellcolor[HTML]{EFEFEF}$\sigma_8\Omega_m^{0.25}$ \\
\hline
MSE $\downarrow$ & 0.001 & 0.01 & 0.0007 & \multicolumn{1}{c}{0.003} & 0.001 & \multicolumn{1}{c|}{0.0008} & 0.0025 & 0.012 & 0.0015 & \multicolumn{1}{c}{0.003} & 0.0015 & 0.001 \\
\cellcolor[HTML]{EFEFEF}$r^2\,\uparrow$ & \cellcolor[HTML]{EFEFEF}0.86 & \cellcolor[HTML]{EFEFEF}0.15 & \cellcolor[HTML]{EFEFEF}0.95 & \multicolumn{1}{c}{\cellcolor[HTML]{EFEFEF}0.04} & \cellcolor[HTML]{EFEFEF}0.83 & \multicolumn{1}{c|}{\cellcolor[HTML]{EFEFEF}0.93} & \cellcolor[HTML]{EFEFEF}0.82 & \cellcolor[HTML]{EFEFEF}0.10 & \cellcolor[HTML]{EFEFEF}0.89 & \multicolumn{1}{c}{\cellcolor[HTML]{EFEFEF}0.05} & \cellcolor[HTML]{EFEFEF}0.75 & \cellcolor[HTML]{EFEFEF}0.92 \\
UCE $\downarrow$ & 0.07 & 0.09 & 0.09 & \multicolumn{1}{c}{\textbf{0.01}} & 0.08 & \multicolumn{1}{c|}{0.20} & \textbf{0.014} & 0.078 & 0.09 & \multicolumn{1}{c}{0.07} & 0.024 & 0.14 \\ \cline{2-13} 
\cellcolor[HTML]{EFEFEF}AV-MSE $\downarrow$ & \multicolumn{6}{c|}{\cellcolor[HTML]{EFEFEF}\textbf{0.0043}} & \multicolumn{6}{c}{\cellcolor[HTML]{EFEFEF}0.0048} \\
NLL $\downarrow$ & \multicolumn{6}{c|}{-95.12} & \multicolumn{6}{c}{-3.34}   \\\hline
\cellcolor[HTML]{EFEFEF}Inf.Time [ms] & \multicolumn{6}{c|}{\cellcolor[HTML]{EFEFEF}345} & \multicolumn{6}{c}{\cellcolor[HTML]{EFEFEF}\textbf{262}}  \\\hline
\end{tabular}
}
\end{table*}
\begin{figure*}
   \resizebox{\hsize}{!}
    {\includegraphics[width=\textwidth]{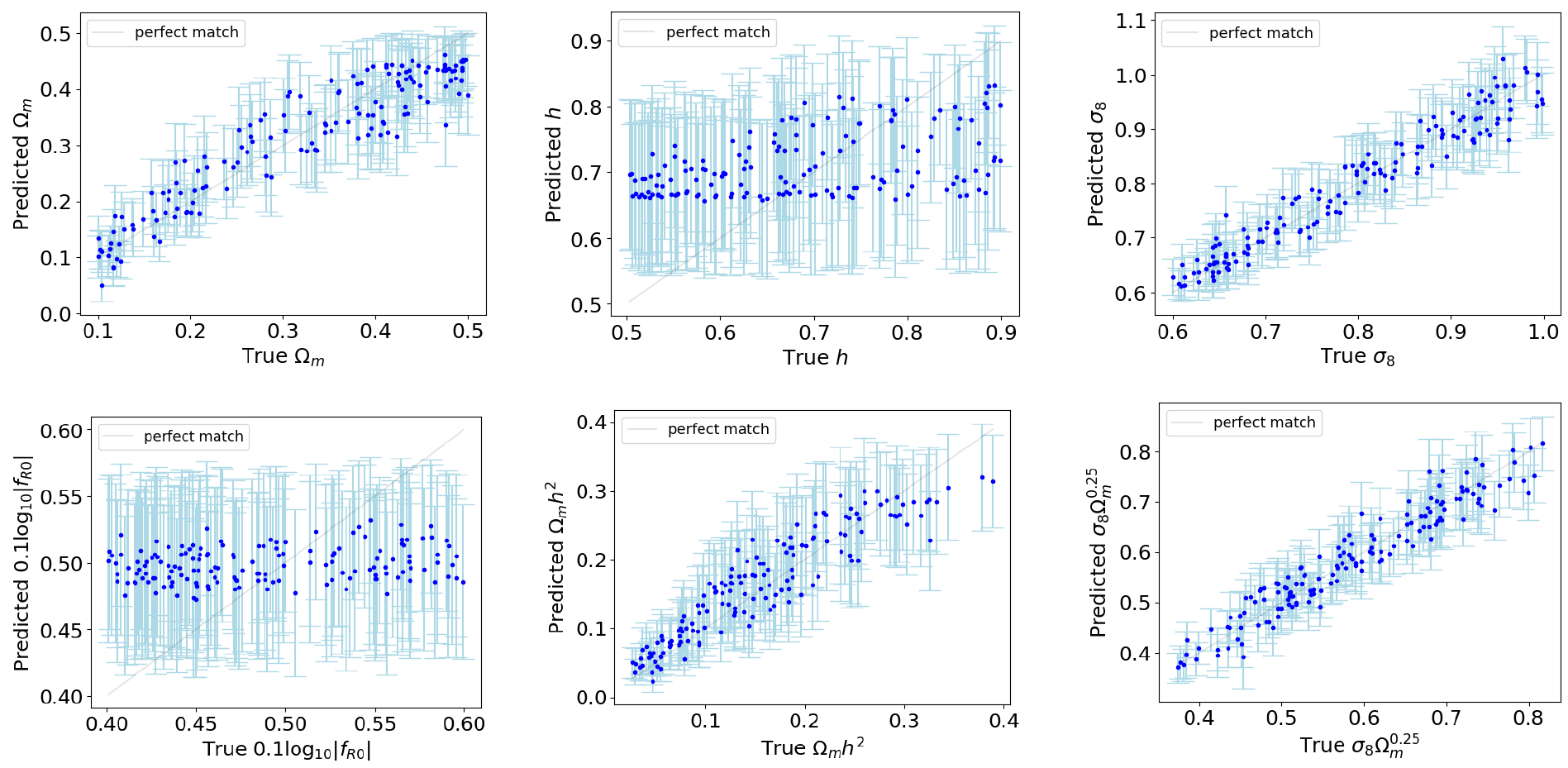}}
    \caption{True vs. predicted values provided by the FullB model for $\Omega_m$, $\sigma_8$, and some derivative parameters. The points are the mean of the predicted distributions, and the error bars stand for the heteroscedastic uncertainty associated with the epistemic and aleatoric uncertainty at $1\sigma$.}
    \label{fig:TruePredicted_VBNNs}
\end{figure*}
The results of the experiments performed in this work are summarised in Table~\ref{tab:dfield_ReSeResNet18a}. It shows the performance of each architecture on the test set. In the top part of the table, the results of the SeResNet18 topology are shown, and in the bottom part, the results of ResNet18  are presented. The left columns of the table correspond to the FullB scheme, and the left colum corresponds to the Bayesian last layer, BLL. Comparing all approaches, we observe that FullB-SeResNet18 slightly outperforms the rest of the models in terms of accuracy (described by $r^2$) and uncertainty quality provided by UCE. However, no significant differences were found in the reported metrics for  ResNet and its SeResNet counterpart, except for the inference time, where the BLL models clearly outperform the FullB models. 
This suggests that FullBs yield small improvements in the computation of the uncertainties at the expense of duplicating the inference time. In addition, both architectures estimate \sigmaeight\ more efficiently than for any other parameter, especially in contrast to $h$ or $0.1\log_{10}|f_{R0}|$, although the FullBs respond slightly better to MG effects. Fig.~\ref{fig:TruePredicted_VBNNs} displays the scatter relation between the predicted and ground-truth values of each cosmological parameter using FullB-SeResNet18.  It also shows the degeneracy directions that arise from observations defined as $\Omega_mh^2$ and $\sigma_8\Omega_m^{0.25}$ (this parameter combination was taken from Planck and CMB lensing~\cite{10.1093/mnras/stad3107}). The diagonal grey lines correspond to the ideal case of a perfect parameter prediction.  Each data point represents the mean value of the predicted distributions, and the error bars stand for the heteroscedastic uncertainty associated with epistemic plus aleatoric uncertainty at 1$\sigma$ confidence level. As we observe, BNNs learn how to accurately predict the value for \Om\ and \sigmaeight, but they fail to capture information related to the MG effects and the Hubble parameter.  Even though parameter estimation derives from all features of the fully nonlinear 3D overdensity field, the horizontal scatter pattern that exhibits the Hubble and MG parameters implies that essential underlying connections are not effectively captured. A similar result for the Hubble parameter using DCNNs in \lcdm\ can be found in \citet{2020ApJS..250....2V}. 

%-------------------------------------------------------------------
\subsection{Parameter estimation from the matter power spectrum}
In this section, we show the results of using the power spectrum to extract the cosmological parameters in MG scenarios. 
\begin{table}[h!]
\caption{Configuration of the fully connected neural network  used for constraining the parameters from the power spectrum.}
\label{tab:BNN_ps}
\small
\centering
\begin{tabular}{ccc}
\hline
 \multicolumn{3}{c}{ \textbf{Fully connected neural network} } \\ [0.1cm]
\cellcolor[HTML]{EFEFEF}\textbf{Layer Name} & \cellcolor[HTML]{EFEFEF}\textbf{Input Shape} & \cellcolor[HTML]{EFEFEF}\textbf{Output Shape} \\ [0.1cm] \hline
Dense Layer & ($N_\text{batch}$, 85) & ($N_\text{batch}$, 64) \\ [0.1cm]
\cellcolor[HTML]{EFEFEF}ReLU & \cellcolor[HTML]{EFEFEF}($N_\text{batch}$, 64) & \cellcolor[HTML]{EFEFEF}($N_\text{batch}$, 64) \\  [0.1cm]
Dense Layer & ($N_\text{batch}$, 64) & ($N_\text{batch}$, 64) \\  [0.1cm]
\cellcolor[HTML]{EFEFEF}  ReLU+Batch Norm & \cellcolor[HTML]{EFEFEF}($N_\text{batch}$, 64) & \cellcolor[HTML]{EFEFEF}($N_\text{batch}$, 64) \\  [0.1cm]
Dense Layer & ($N_\text{batch}$, 64 ) & ($N_\text{batch}$, 64) \\ [0.1cm]
\cellcolor[HTML]{EFEFEF} ReLU & \cellcolor[HTML]{EFEFEF}($N_\text{batch}$,  64)  & \cellcolor[HTML]{EFEFEF}($N_\text{batch}$,  64)\\ [0.1cm]
Dense Layer & ($N_\text{batch}$, 64 ) & ($N_\text{batch}$, 14) \\ [0.1cm]
 \cellcolor[HTML]{EFEFEF} Multivariate normal & \cellcolor[HTML]{EFEFEF}($N_\text{batch}$,  14) & \cellcolor[HTML]{EFEFEF}($N_\text{batch}$,  14) \\ 
\hline
\end{tabular}
\end{table}
\begin{table*}[]
\caption{Metrics for the power spectra test set with fully connected networks (FCN). High UCE values indicate miscalibration.  MSE and NLL are computed only over the cosmological parameters. The bold text is the minimum (maximum) value, $\downarrow$ ($\uparrow$) as indicated  in the metric name, among the different parameters.}\label{table2ps}
\resizebox{\linewidth}{!}{%
\begin{tabular}{l cccccc cccccc}
\hline
\multirow{2}{*}{\textbf{Metrics}} & \multicolumn{6}{c|}{\textbf{FullB-FCN }} & \multicolumn{6}{c}{\textbf{BLL-FCN}} \\

         & \cellcolor[HTML]{EFEFEF}$\Omega_m$ & \cellcolor[HTML]{EFEFEF}$h$ & \cellcolor[HTML]{EFEFEF}$\sigma_8$ & \cellcolor[HTML]{EFEFEF}$0.1\log_{10}|f_{R0}|$ & \cellcolor[HTML]{EFEFEF}$\Omega_mh^2$ & \multicolumn{1}{c|}{\cellcolor[HTML]{EFEFEF}$\sigma_8\Omega_m^{0.25}$} & \cellcolor[HTML]{EFEFEF}$\Omega_m$ & \cellcolor[HTML]{EFEFEF}$h$ & \cellcolor[HTML]{EFEFEF}$\sigma_8$ & \cellcolor[HTML]{EFEFEF}$0.1\log_{10}|f_{R0}|$ & \cellcolor[HTML]{EFEFEF}$\Omega_mh^2$ & \cellcolor[HTML]{EFEFEF}$\sigma_8\Omega_m^{0.25}$ \\
\hline
MSE $\downarrow$ & 0.0023 & 0.012 & 0.0007 & \multicolumn{1}{c}{0.003} & 0.0013 & \multicolumn{1}{c|}{0.0011} & 0.0023 & 0.011 & 0.00078 & \multicolumn{1}{c}{0.0030} & 0.0012 & 0.0012 \\
\cellcolor[HTML]{EFEFEF}$r^2\,\uparrow$ & \cellcolor[HTML]{EFEFEF}0.83 & \cellcolor[HTML]{EFEFEF}0.11 & \cellcolor[HTML]{EFEFEF}0.94 & \multicolumn{1}{c}{\cellcolor[HTML]{EFEFEF}0.06} & \cellcolor[HTML]{EFEFEF}0.77 & \multicolumn{1}{c|}{\cellcolor[HTML]{EFEFEF}0.90} & \cellcolor[HTML]{EFEFEF}0.83 & \cellcolor[HTML]{EFEFEF}0.16 & \cellcolor[HTML]{EFEFEF}0.94 & \multicolumn{1}{c}{\cellcolor[HTML]{EFEFEF}0.073} & \cellcolor[HTML]{EFEFEF}0.80 & \cellcolor[HTML]{EFEFEF}0.89 \\
UCE $\downarrow$ & 0.026 & 0.12 & 0.022& \multicolumn{1}{c}{\textbf{0.022}} & 0.026 & \multicolumn{1}{c|}{\textbf{0.092}} & \textbf{0.023} & \textbf{0.15} & \textbf{0.017} & \multicolumn{1}{c}{0.023} & \textbf{0.016} & 0.10 \\ \cline{2-13} 
\cellcolor[HTML]{EFEFEF}AV-MSE $\downarrow$ & \multicolumn{6}{c|}{\cellcolor[HTML]{EFEFEF}0.0045} & \multicolumn{6}{c}{\cellcolor[HTML]{EFEFEF}\textbf{0.0043}} \\
NLL $\downarrow$ & \multicolumn{6}{c|}{64.86} & \multicolumn{6}{c}{1.80} \\\hline
\cellcolor[HTML]{EFEFEF}Inf.Time [ms] & \multicolumn{6}{c|}{\cellcolor[HTML]{EFEFEF}3.01} & \multicolumn{6}{c}{\cellcolor[HTML]{EFEFEF}\textbf{2.21}}  \\\hline
\end{tabular}
}
\end{table*}
Following the same method as described in the voxel-grid representation, we implemented two BNN models that provided distributed predictions for the cosmological parameters. Table~\ref{tab:BNN_ps} schematically presents the architecture used for this purpose. This represents a fully connected network (FCN) with 60000 trainable parameters, and it was derived from KerasTuner\footnote{\url{https://keras.io/keras_tuner/}}  as a framework for a scalable hyperparameter optimisation. We worked with a Bayesian last layer model (BLL-FCN) along  with a full Bayesian topology where all dense layers are probabilistic (FullB-FCN). Here, the power spectrum computed from the N-body simulations was kept until $k \approx 1.58$\Mpch, obtaining arrays of 85 dimensions. The results of this approach are shown in Table~\ref{table2ps}. In contrast to the voxel-grid representation, where the full Bayesian approach outperforms most of the models, here we clearly observe that the BLL approach works better than the fully Bayesian approach.  These results show a similar performance as the 3D overdensity field. We expected this behaviour since most of the voxel-grid information should be encoded in the two-point correlator. Furthermore, some parameters such as $\sigma_8$ or the derived parameters provide a higher accuracy when they are predicted with the voxel-grid approach, supporting the fact that a 3D convolutional layer extracts further information beyond the linear part.
The interplay between the $f_{R0}$ parameter and the shape of the power spectrum is essential for testing and constraining gravity theories. The immediate effect of $f_{R0}$ on the power spectrum is to modulate its amplitude, most notably, at small scales. Furthermore, this parameter of the HS model exhibits a substantial degeneracy with \sigmaeight, which produces a similar effect on the power amplitude, but not in a scale-dependent manner, as MG does. The strongest deviations of the power spectrum from the \lcdm\ model are observed for high values of $f_{R0}$, in our case, $\sim10^{-4}$ (see \cref{fig:pks_sim}). Because of this degeneracy, it is probable that some of the MG information is encoded in the \sigmaeight\ parameter rather than the $f_{R0}$ parameter. This hypothesis, however, would require additional tests of the BNN with a reduced parameter space in addition to isolating the impact of the sole case of a zero $f_{R0}$, which we leave for future work.

%-------------------------------------------------------------------
\subsection{Comparison of the approaches  based on marginalised parameter constraints}
\begin{figure*}
   \centering
   \includegraphics[width=0.7\textwidth]{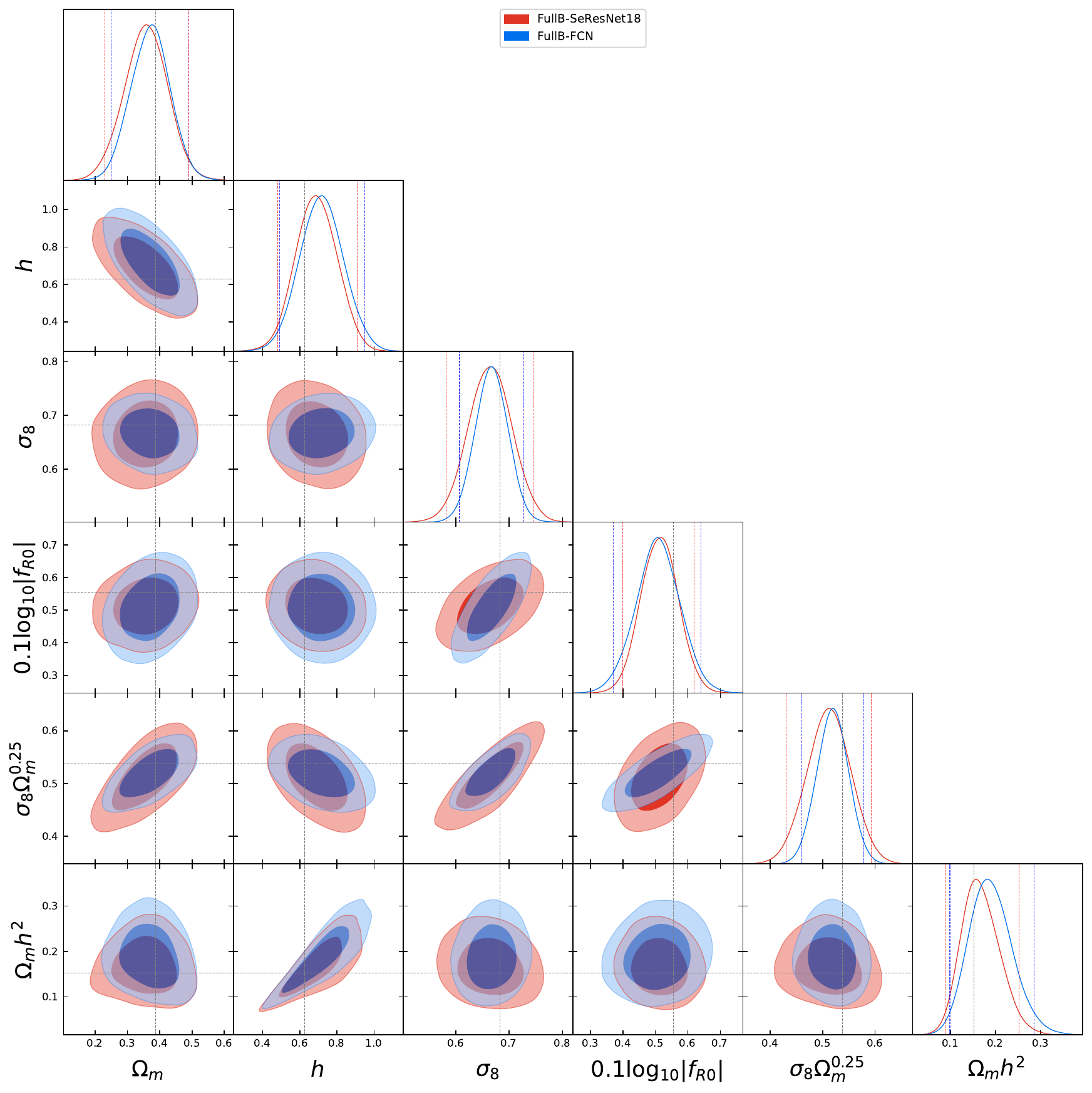}
      \caption{$68\%$ and $95\%$ parameter constraint contours from one example of the test dataset using FullB-SeResNet and Full-FCN. The diagonal plots are the marginalised parameter constraints, and the dashed black lines stand for the true values reported in Table~\ref{KapSou}. The vertical dashed red and blue lines represent the $1\sigma$ for the marginals using FullB-SeResNet and Full-FCN, respectively.   We derived these posterior distributions using GetDist~\citep{Lewis:2019xzd}.
              }
         \label{fig:triangular-plot}
   \end{figure*}

Finally, we chose one example from the test set to compare the constrain contours predicted by the best models presented in the paper so far. Fig.~\ref{fig:triangular-plot} compares the parameter constraints at $68\%$ and $95\%$  confidence levels predicted for the FullB-SeResNet18 and FullB-FCN models.   
The true values of the example are reported in  Table~\ref {KapSou} and are represented by dashed lines in the triangular plot. 
\begin{table}
\centering
      \caption[]{Parameters in the 95\% intervals taken from the parameter constraint contours from one example of the MG simulations test set predicted by the FullB-SeResnet18 and FullB-FCN.}
         \label{KapSou}
\begin{tabular} { l  c c c }
\hline
 Parameter &  SeResNet18  &    FCN & Target\\
\hline
\cellcolor[HTML]{EFEFEF}{\boldmath$\Omega_m       $} & \cellcolor[HTML]{EFEFEF}$ 0.36^{+0.13}_{-0.13}   $ & \cellcolor[HTML]{EFEFEF} $ 0.37^{+0.12}_{-0.12}$ & \cellcolor[HTML]{EFEFEF} $0.3865$ \\

{\boldmath$h              $} & $  0.69^{+0.22}_{-0.21}    $ & $   0.72^{+0.23}_{-0.23}    $& $  0.6274   $\\

\cellcolor[HTML]{EFEFEF}{\boldmath$\sigma_8       $} & \cellcolor[HTML]{EFEFEF}$ 0.664^{+0.081}_{-0.082} $ &  \cellcolor[HTML]{EFEFEF} $ 0.667^{+0.060}_{-0.060}  $ & \cellcolor[HTML]{EFEFEF} $0.6822$ \\

{\boldmath$   0.1\log_{10}|f_{R0}|    $} & $   0.51^{+0.11}_{-0.11}  $ &  $  0.51^{+0.13}_{-0.14}    $& $  0.5557  $\\

\cellcolor[HTML]{EFEFEF}$\sigma_8\Omega_m^{0.25}   $ & \cellcolor[HTML]{EFEFEF}$ 0.512^{+0.081}_{-0.082}  $  & \cellcolor[HTML]{EFEFEF} $ 0.519^{+0.059}_{-0.059} $ & \cellcolor[HTML]{EFEFEF} $0.5379$ \\

$\Omega_m h^{2}            $ & $  0.167^{+0.085}_{-0.079}   $  & $   0.190^{+0.096}_{-0.091}    $ & $ 0.1521  $\\
\hline
\end{tabular}
\end{table}
Both models yield decent predictions for the marginal distribution, but they differ in the correlation of the cosmological parameters, as  \sigmaeight\ and $f_{R0}$, where this behaviour is more notorious. This clearly implies that 3D convolutions extract further information beyond the linear regime that allows us to constrain the parameter estimation more tightly.  
\begin{table}[]
    \caption[]{Relative error comparison of the different CNN approaches for MG and \lcdm\ simulations. The relative error has been defined as $\delta y\equiv\Delta y/y$, where $y$ stands for \Om, \sigmaeight\, and $\Delta y$ is the uncertainty.}
    \label{tab:relerrorcomp}
    \begin{tabular}{lccl}
    \hline
    Method             & $\delta\Omega_m$    & $\delta \sigma_8$ & Reference      \\\hline
    \rowcolor[HTML]{EFEFEF} 
    CNN                & 0.0048 & 0.0053  & \citet{2020SCPMA..6310412P}      \\
    CNN                & 0.0280 & 0.0120  & \citet{2017arXiv171102033R}    \\
    \rowcolor[HTML]{EFEFEF} 
    VBNNs              & 0.2128 & 0.0545  & \citet{frontiershector}   \\
    FlipoutBNN         & 0.2444 & 0.0844  & \citet{frontiershector}   \\
    \rowcolor[HTML]{EFEFEF} 
    SeResNet & 0.3611 & 0.1220  & This work \\
    FCN        & 0.3243 & 0.0900  & This work \\\hline
    \end{tabular}
    \end{table}

%-----------------------------------------------------------------
\section{Summary and discussion}
We considered a wide range of MG simulations for which we varied the cosmological parameters. They encompassed cosmologies with large deviations from the standard GR to parameters closest to those that mimic the dynamics of a Universe based on GR. The overdensity field of each snapshot was computed using the CIC mass assignment, and subsequently, we obtained its power spectrum. To constrain the main set of cosmological parameters, we introduced a novel architecture of a BNN and designed several experiments to test its ability to predict MG cosmologies. The experiments consist of building two Bayesian networks based on stochastic layers located at either the top or at all levels of the architecture. This approach was motivated by the question of whether BNNs provide a better accuracy and robustness performance when we work with full or partial network configurations. Starting from the 3D overdensity field, we found that although the FullB predicts the cosmological parameters slightly better than the BLL, the latter is accurate enough to retrieve cosmological information from the density field, especially for \Om\ and \sigmaeight. Similarly, we tested  BNNs using the two-point statistics described by the power spectrum for reasonable scales limited by the Nyquist frequency. The results of this experiment show that the information learned by the networks can predict the parameters with an accuracy similar to the 3D field. Both configurations of the BNN architectures fall short of capturing both the Hubble parameter and the MG effects. This underscores the necessity of improving the training dataset in terms of resolution and scale for the 3D density setup.  Despite the slight constraints for some cosmological parameters, the method can be relevant in applications where it is combined with classical inference methods~\citep{Hort_a_2020}. The multiplicative normalising flows technique in BNNs employed in this paper has proved to yield good predictions and accurate uncertainty estimates through the ability to transform the approximate posterior into a more expressive distribution consistent with the data complexity. This is a significant improvement compared to standard VI, where the posterior is restricted to a Gaussian configuration. Nevertheless, the effect of assuming a Gaussian prior distribution of the weights under this approach is still unknown~\citep{fortuin2022bayesian}. In future work, we will explore multiplicative normalising flows with different prior distributions over the weights and analyse how the prior influences the uncertainty calibration and performance. 

The finding that the MG parameter is poorly predicted when the information provided by the density field is used demonstrates, on the one hand, the effectiveness of the chameleon screening mechanism in mimicking the \lcdm\ model, and on the other hand, the need for further analysis with other datasets that are more sensitive to the effects of MG. We considered parameters that produce the same effect, that is, that are degenerate. Therefore, it is not straightforward to attribute a single characteristic of the overdensity field exclusively to a single parameter, as in the case of $f_{R0}$ and \sigmaeight.   The proposed architectures are sufficiently general from a statistical standpoint to estimate posterior distributions. However, this study has revealed that the available information is inadequate to predict all parameters solely from a single source. This underscores the significance of resolving degeneracies between cosmological parameters by incorporating supplementary data or diverse features present in the cosmological simulations. This approach enables the BNNs to gain a richer learning phase and parse out the signals of each cosmology. This task will be the focus of a forthcoming paper, where we plan to evaluate the BNN robustness using simulations of higher resolution and more intricate datasets in redshift space, incorporating velocity information alongside particle positions.

In Table \ref{tab:relerrorcomp}, we also presented a comparison of the relative errors for the two best-estimated parameters using CNN and N-body simulations from the literature. We observed significant discrepancies in the relative errors of \sigmaeight\ and \Om, approximately 90\% when Bayesian inference is not employed \citep[see][]{2017arXiv171102033R, 2020SCPMA..6310412P}. This outcome arises from using solely \lcdm\ simulations in both training and test datasets, in contrast to our estimates, which encompass an additional parameter accounting for MG and include a calibration procedure of the uncertainties. Furthermore, when comparing the performance of BLL architectures on MG and \lcdm\ simulations, such as \texttt{QUIJOTE} \citep[see \eg][]{frontiershector}, we find a deviation of the relative errors close to 30\% when MG effects are not considered. This result clarifies that when using FullB-SeResNet18, the error bars for \Om\ are 1.3 times larger and are larger by 2.1 times for \sigmaeight\  in comparison to FlipoutBNN. In the context of BNNs, when separately considering the two cosmological models MG and \lcdm\, we assessed the performance in terms of the MSE metric by comparing it to the results presented by \citet{frontiershector}, who employed a similar architecture. Specifically, using FullBs in both cosmological models, we observed an improvement by a factor 13 in the MSE of the MG predictions over the \lcdm\ ones. The $r^2$ metric was used to compare the confidence range of the individual parameters. In terms of this metric, we report that \sigmaeight\ deviates more ($r^2$ = 0.95 in MG and $r^2=$0.99 in \lcdm), which accounts for 4.2\% of the expected uncertainty. The marginal difference in the coefficient of determination for predicting \Om\ is only 0.01 when comparing the results of the model trained with MG against the model trained with \lcdm. In both cases, it is noteworthy that a high $r^2$ value does not necessarily confer complete certainty regarding individual parameter estimates, particularly when the parameter degeneracy is taken into account. Furthermore, one interesting possibility to refine the constraints on $f(R)$ gravity is given by training a specialised network that clearly distinguishes between \lcdm\ and $f(R)$, offering the potential to detect a non-zero $f_{R0}$. Further investigations, including high-resolution simulations as well as extensions beyond \lcdm, promise to further enhance the capabilities of the BNNs approach. 
The BNNs prove valuable in constructing mock catalogues for galaxy surveys, excelling in tasks such as object classification and feature extraction. The synergy of BNNs with galaxy painting techniques further strengthens the ability to capture complex patterns in the data, offering valuable insights into large-scale studies. The BNN model we developed can be readily adapted for the analysis of observational data. However, this modification requires a more substantial effort in appropriately processing the inputs to align with the noise level inherent in dedicated galaxy surveys. Addressing challenges such as bias, sample tracers, peculiar motions, and the systematic from galaxy surveys becomes imperative in this context. They might nonetheless be implemented in Bayesian inference algorithms of the large-scale structure \citep[such as, e.g.,][]{2021MNRAS.502.3456K}.  Together with this paper, we make the scripts available, which can be accessed at the public github repository of the project\footnote{\url{https://github.com/JavierOrjuela/Bayesian-Neural-Net-with-MNFs-for-f-R-}}.

%-------------------------------------------------------------------
\section{Conclusions}
One of the intriguing possibilities to explain the observed accelerated expansion of the Universe is the modification of general relativity on large scales. Matter distribution analysis via N-body simulations offers a perfect scenario to track departures from standard gravity. Among different parametrisations, $f(R)$ has emerged as an interesting model because it can reproduce the standard model predictions accurately. In this paper, we analysed the possibility of using Bayesian deep-learning methods for constraining cosmological parameters from MG simulations.  Below, we summarise the main takeaways from this study.
 \begin{enumerate}
      \item BNNs can predict cosmological parameters with a higher accuracy, especially for $\Omega_m$ and $\sigma_8$ from the overdensity field. However, based on the assumption of simulating boxes with 256 \Mpch\ to acquire MG effects on large scales, BNNs were unable to effectively extract MG patterns from the overdensity field to yield an accurate $f(R)$ parameter estimation. However, when comparing the parameter estimation with \lcdm-only simulations, the uncertainties of \sigmaeight\ are significantly underpredicted when possible MG effects are not taken into account.
      In addition, special attention should be paid to parameter degeneracies that may be present not only in two-point statistics, but in more features of the density field. We conclude that higher resolution and further intricate datasets in redshift space, incorporating velocity information alongside particle positions, can be approaches that should be addressed to improve the network predictions. 
      \item It is observed that cosmological parameters can be recovered directly from the simulations using convolution-based models with the potential of extracting patterns without specifying any N-point statistics beforehand. This is supported by the fact that networks trained with overdensity fields and power spectra yielded decent predictions, but with distinctive correlations among the parameters. 3D convolutions extracted supplementary information beyond the linear regime, which allowed them to constrain the parameter estimation tightly. 
      \item We generalised the multiplicative normalising flows for BNNs to the 3D convolutional level, which allowed us to work with fully transformed stochastic neural networks.  As a proof of concept, we ran several experiments to verify that this approach not only achieved the performance reached by the deterministic models, but also yielded well-calibrated uncertainty estimates. 
      \item We probed the impact of the parameter estimation based on the  Bayesian last layer (BLL) and fully Bayesian approaches. The results showed that fullBs provide slightly higher-quality predictions along with accurate uncertainty estimates. Nevertheless,  this improvement is not significant enough to prefer this approach to the BLL, where the latter has the advantage of being relatively model agnostic, easily scalable, and is twice as fast in inference time. 
      \item Several experiments have reported that normalising flows added at the output layers as well to avoid bottlenecks by the simple multivariate distribution do not improve the model performance significantly. We therefore decided to work with the simple distribution as output for the network.
   \end{enumerate}

% -----------------------------------------------
\begin{acknowledgements}
This paper is based on work supported by the Google Cloud Research Credits program with the award GCP19980904. HH acknowledges support from créditos educación de doctorados nacionales y en el exterior-Colciencias and the grant provided by the Google Cloud Research Credits program.\\
JEGF is supported by the Spanish Ministry of Universities, through a Mar\'ia Zambrano grant (program 2021-2023) at Universidad de La Laguna with reference UP2021-022, funded within the European Union-Next Generation EU. FSK and JEGF acknowledge the IAC facilities and the Spanish Ministry of Science and Innovation (MICINMINECO) under project PID2020-120612GB-I00.  We also thank the personnel of the Servicios Inform\'aticos Comunes (SIC) of the IAC.
\end{acknowledgements}

\bibliographystyle{aa}
\bibliography{bibliography} 
\end{document}